\documentclass[a4paper,12pt]{article}
\usepackage[dvips]{graphicx}
\usepackage[cp 1251]{inputenc}
\begin{document}

\begin{center}
{\large \textbf{ Momentum distribution of $\Delta$- isobar in closed shell
 nuclei.}}
 \end{center}
 \begin{center}
 A. N. Tabachenko\\
 \end{center}
 \begin{center}
 Nuclear Physics Institute , 634050 Tomsk,  Russia
 \end{center}
 \begin{center}
One $\Delta$- isobar components of the wave function in closed
shell nuclei are considered within the framework of the harmonic
oscillator model. Conventional transition potential is the  $\pi
$- and $\rho$- exchange potential. On the basis of the $ \Delta$-
isobar configuration wave function, the momentum distribution of
the $\Delta$- isobar is calculated for the light nuclei $^4
He$,$^{16}O$,$^{12}C$.
\end{center}
            \begin{center}
  \textbf{Introduction}
            \end{center}
Within the framework of the traditional non-relativistic theory of
nuclei, the non-nucleon degrees of freedom are bound up with the
effect of mutual polarization or deformation of bound nucleons in
nuclei  due to close  collisions within nuclei [1- 4]. In terms of
the wave functions of  nuclei this effect can be expressed by the
introduction of additional to purely nucleon, isobar
configurations. These configurations describe such states of
nuclei, in which the part of nucleons in nuclei are in excited
states, as virtual isobars, which appear due to the collisions
within nuclei as a result the transitions
   $$
   N+ N\rightarrow N+\Delta (\Delta+\Delta)\rightarrow N+N .
   $$
The admixture of probability of these exotic nuclear configurations is small because
of the small nuclear density and big isobar-nucleon mass difference. These exotic
$\Delta$- isobars (internal $\Delta$- isobars) are far off the mass-shell unlike
$\Delta$- isobars (external $\Delta$- isobars), which are born in reactions of
particles with nuclei and are essentially on the mass-shell [5]. There were attempts
to find  manifestations of the exotic $\Delta$- isobar configurations in the ground
state of light nuclei in a number of experiments, in reactions on nuclei with
knocking out $\Delta$-isobars preexisting in the target nuclei[6-10]. A theoretical
description of these reactions within the framework of the impulse approximation
requires knowledge of the momentum distribution of $\Delta$- isobars considered as
components of nucleus, together with nucleons. This momentum distribution differs
from the momentum distribution of the external isobars and does not depend on
kinematics and characteristics of particles in reactions with nuclei.

In this work, the estimation is made for the probability of the
internal $\Delta$- isobars presence in light nuclei, and the
momentum distribution of the $\Delta$- isobars in light nuclei
with the closed shells is obtained.
    \begin{center}
 \textbf{ Wave functions of isobar configurations in closed shell nuclei.}
   \end{center}
According to the approach developed in the works of Arenh$\ddot{o}$vel et al.
[5,11,12], nucleons bound in the nucleus, in addition to the spatial, spin, and
isospin coordinates, are characterized also by the intrinsic coordinate.  For
completeness and to fix the notation, let us summarize the results and formulas
given in [11]. The state vector of N nucleons is $$
\mid\alpha_{1}(1),...\alpha_{N}(N)>=\mid\beta_{1}(1)n_{1}(1),...\beta_{N}(N)n_{N}(N)>.
$$ The indices $\beta_{1},...\beta_{N}$ refer to the usual,
spin, and isospin space. They take the same series of values for all particles,
i.e., $\beta_{1}= \alpha,\beta,\gamma...;...\beta_{N}=\alpha,\beta,\gamma...$. The
indices $ n_{1}(1),...n_{N}(N)$ characterize the state vectors in the intrinsic
space.
 The basis vectors $\mid m_{\nu}>$ in the intrinsic space are defined as follows
$$
 \hat{M}\mid m_{\nu}>=m_{\nu} \mid m_{\nu}>,
$$
 where $m_{\nu}$ is the eigenvalue of mass operator $\hat{M}$ in the intrinsic space, and
$\nu$ =1, 2, 3,.... The value of $m_{\nu=1}$ corresponds to a mass of nucleon,
$m_{\nu=2}$ corresponds to
 a mass of first excited state - $\Delta$, and so on. The vectors $\mid m_{\nu}>$ form
the infinite discrete basis in the intrinsic space. Each particle is characterized
by its intrinsic  index of state $n_{i}(i) (i=1...N)$. Later on, it is taken that
$\mid n_{i}(i)>$=$\mid m_{\nu=1}>$, $\mid m_{\nu=2}>$,...., i.e., states are pure
states. For brevity, we enter the designation $\mid m_{\nu=1}>$=$\mid N>$, $\mid
m_{\nu=2}>$=$\mid \Delta>$,.... For instance, the vector describing the system of N
nucleons is written as $\mid N(1)...N(N)>$; if the first particle is in the state of
isobar, but the rest are nucleons, the state vector  is written as $\mid
\Delta(1)...N(N)>$.

For the system of N particles, which can be in different intrinsic states,
hamiltonian of system H acts on spatial, spin, isospin, and intrinsic coordinates.
According to [11] hamiltonian H  have form
$$
  H=\sum_{k=1}^{A}(T(k)+H_{in}(k))+\sum_{i\neq k}V(i,k).
  $$
 Here, T(k) is the kinetic energy operator of k- particle, $H_{in}(k)$ is the part
connected with the intrinsic degrees of freedom, V(i,k) is the
two-particle interaction. The operators T and V, unlike those of
standard nuclear physics, depend also on the intrinsic degrees of
freedom. The operators T and $H_{in}$ are diagonal on the
intrinsic degrees of freedom. Let us assume that $$
H_{in}=\hat{M}-\hat{I}M_{N}, $$ i.e., the vectors $\mid m_{\nu}>$
are eigenvectors of the operator $H_{in}$ with the eigenvalue
$(m_{\nu}-M_{N})$.
 Thus, the following formulae take place
 $$
  <m_{\nu^{'}}\mid H_{in}\mid
m_{\nu}>=(m_{\nu^{'}}-M_{N})\delta_{\nu^{'}\nu},
 $$
  $$
 <m_{\nu^{'}}\mid T\mid
m_{\nu}>=\frac{p^{2}}{2M_{\nu}}\delta_{\nu^{'}\nu},
$$
 $$
<m_{\nu^{'}}m_{\mu^{'}}\mid V \mid m_{\nu}m_{\mu}>= V_{m_{\nu^{'}}m_{\mu^{'}},
m_{\nu}m_{\mu}}.
$$
The wave function of the system of N particles in the state
 $\mid\alpha_{1}(1),...\alpha_{N}(N)>$ characterized also by the
 intrinsic coordinates is introduced as follows
 $$
\psi_{\alpha_{1},...\alpha_{N}}(\vec{r}_{1},\sigma_{z}(1),\tau_{z}(1),m_{\nu_{1}},...
\vec{r}_{N},\sigma_{z}(N),\tau_{z}(N),m_{\nu_{N}})= $$ $$
<\vec{r}_{1},\sigma_{z}(1),\tau_{z}(1),m_{\nu 1},...
\vec{r}_{N},\sigma_{z}(N),\tau_{z}(N),m_{\nu_{N}}\mid\alpha_{1}(1),...\alpha_{N}(N)>.
$$
 An eigenfunction $\Psi_{\beta_{1},...\beta_{N}}(\vec{r}_{1},...
\vec{r}_{N};m_{\nu{1}}^{\prime}...m_{\nu_{N}}^{\prime})$ of N
particles of operator H with eigenvalues
$E_{\beta_{1},...\beta_{N}}$ is a superposition of the wave
functions belonging to the different configurations
 $$ \Psi_{\beta_{1},...\beta_{N}}(\vec{r}_{1},...
\vec{r}_{N};m_{\nu{1}}^{\prime}...m_{\nu_{N}}^{\prime})=
\sum_{n_{1}^{\prime},...n_{N}^{\prime}}
A_{n_{1}^{\prime},...n_{N}^{\prime}}\phi_{n_{1}^{\prime},...n_{N}^{\prime}}
(m_{\nu{1}}^{\prime}...m_{\nu_{N}}^{\prime}) \cdot
\psi_{\beta_{1},... \beta_{N}}^{n_{1}^{\prime},...n^{\prime}_{N}}
(\vec{r}_{1},... \vec{r}_{N}).
$$
 For brevity, we enter the
designation $\vec{r}$ for space, spin, and isospin coordinates.
 The part of the wave function of nucleus
$A_{n_{1}^{\prime},...n_{N}^{\prime}}\phi_{n_{1}^{\prime},...n_{N}^{\prime}}
(m_{\nu{1}}^{\prime}...m_{\nu_{N}}^{\prime}) \cdot \psi_{\beta_{1},...
\beta_{N}}^{n_{1}^{\prime},...n^{\prime}_{N}} (\vec{r}_{1},... \vec{r}_{N})$
describes configuration of N particles in nucleus with quantum numbers of the
intrinsic states $ n'_{1},...n'_{N}$. For instance, the nucleon part of the nuclear
wave function is characterized by the values $n'_{1}=N(1),...,n'_{N}=N(N)$,
one-isobar configuration have $n'_{1}=\Delta(1),...,n'_{N}=N(N)$ and so on. By
definition, $\psi_{\beta_{1},...\beta_{N}}^{n'_{1},...n'_{N}} (\vec{r}_{1}, ...
\vec{r}_{N})$ is the wave function of space, spin, and isospin coordinates of the
configuration of N particles with quantum numbers $ n'_{1},...n'_{N}$. The wave
function $ \phi_{n'_{1},...n'_{N}} (m_{\nu{1}}...m_{\nu_{N}})=
<m_{\nu_{1}}(1)...m_{\nu_{N}}(N)\mid n'_{1}(1)...n'_{N}(N)>$ is the intrinsic wave
function of N particles. The wave function
$\psi_{\beta_{1},...\beta_{N}}^{n'_{1},...n'_{N}} (\vec{r}_{1}, ... \vec{r}_{N})$
should be antisymmetric for particles in the same intrinsic state. The remaining
antisymmetrization for particles in different intrinsic states is done by operator
$A_{n_{1}^{\prime},...n_{N}^{\prime}}$.

The wave functions
$A_{n_{1}^{\prime},...n_{N}^{\prime}}\phi_{n_{1}^{\prime},...n_{N}^{\prime}}
(m_{\nu{1}}^{\prime}...m_{\nu_{N}}^{\prime}) \cdot \psi_{\beta_{1},...
\beta_{N}}^{n_{1}^{\prime},...n^{\prime}_{N}} (\vec{r}_{1},... \vec{r}_{N})$
 satisfy the following Schr$\ddot{o}$dinger equation
$$
 \sum_{m_{\nu_{1}}...m_{\nu_{N}
 }; m_{\nu^{'}_{1}}...m_{\nu^{'}_{N}}}
 \phi_{n_{1}...n_{N}}( m_{\nu_{1}}...m_{\nu_{N}})
(H(\vec{r_{1}}...\vec{r_{N}})-E_{\beta_{1},...\beta_{N}})_{m_{\nu_{1}}...m_{\nu_{N}
 }; m_{\nu^{'}_{1}}...m_{\nu^{'}_{N}}}
 $$
 $$
 A_{n_{1},...n_{N}} \phi_{n_{1},...n_{N}}(m_{\nu{1}}^{\prime}...m_{\nu_{N}}^{\prime}) \cdot
 \psi_{\beta_{1},...\beta_{N}}^{n_{1},...n_{N}}
(\vec{r}_{1},... \vec{r}_{N})=
$$
$$
-\sum_{m_{\nu_{1}}...m_{\nu_{N}
 }; m_{\nu^{'}_{1}}...m_{\nu^{'}_{N}}}
 \phi_{n_{1}...n_{N}}( m_{\nu_{1}}...m_{\nu_{N}})
(V)_{m_{\nu_{1}}...m_{\nu_{N}
 }; m_{\nu^{'}_{1}}...m_{\nu^{'}_{N}}}
 $$
$$ \sum_{n_{1}^{\prime},...n_{N}^{\prime}\neq
n_{1},...n_{N}}A_{n_{1}^{\prime},... n_{N}^{\prime}}
\phi_{n_{1}^{\prime},...
n_{N}^{\prime}}(m_{\nu{1}}^{\prime}...m_{\nu_{N}}^{\prime}) \cdot
\psi_{\beta_{1},...\beta_{N}}^{n_{1}^{\prime},...n^{\prime}_{N}}
(\vec{r}_{1},... \vec{r}_{N}). $$

 There is no hope of solving this
set of equations, because of the right- hand side of this equation couples the
various isobar configurations. In practice, the impulse approximation is used,
according to which on the right- hand side of this equation
 only the terms with the wave functions of nucleon
configuration are leaved. Besides, the interactions between isobars and between
isobars themselves are neglected on the left- hand side of equation. This
approximation leaves only one-isobar and two- isobar configurations.

The full wave function
$\Psi_{\beta_{1},...\beta_{N}}(\vec{r}_{1},...\vec{r}_{N};
m_{\nu{1}}^{\prime}...m_{\nu_{N}}^{\prime})$ must be normalized to
the unit. Let us consider a bilinear form
$$
 \int d1...dN {\Psi^{\ast}}_{\beta_{1},...\beta_{N}}(\vec{r}_{1},...
\vec{r}_{N};m_{\nu{1}}^{\prime}...m_{\nu_{N}}^{\prime})
\Psi_{\beta_{1},...\beta_{N}}(\vec{r}_{1},...
\vec{r}_{N};m_{\nu{1}}^{\prime}...m_{\nu_{N}}^{\prime}) $$ Here,
d1...dN means an integral of the space variable and the summation
over the spin, isospin, and intrinsic coordinates. Using the
condition of orthogonality for the intrinsic wave functions
 $$
\sum_{m_{\nu{1}}^{\prime}...m_{\nu_{N}}^{\prime}}\phi^{\ast}_{n_{1}
,...n_{N}}
(m_{\nu{1}}^{\prime}...m_{\nu_{N}}^{\prime})\phi_{n_{1}^{\prime},...n_{N}^{\prime}}
(m_{\nu{1}}^{\prime}...m_{\nu_{N}}^{\prime})=
\delta_{n_{1}n_{1}^{\prime}}...\delta_{n_{N}n_{N}^{\prime}}
 $$
  and $A^{+}=A$, $A^{2}=\sqrt{N!}A $, we obtain
$$
\int d1...dN {\Psi^{\ast}}_{\beta_{1},...\beta_{N}}(\vec{r}_{1},...
\vec{r}_{N};m_{\nu{1}}^{\prime}...m_{\nu_{N}}^{\prime})
\Psi_{\beta_{1},...\beta_{N}}(\vec{r}_{1},...
\vec{r}_{N};m_{\nu{1}}^{\prime}...m_{\nu_{N}}^{\prime})=
  $$
  $$
 \int d\vec{r_{1}}...d\vec{r_{N}}
 \sum_{n_{1}^{\prime},...n^{\prime}_{N}} {\psi^{\ast}}_{\beta_{1},...
 \beta_{N}}^{{n_{1}^{\prime}},{n_{1}^{\prime}}
 ...{n^{\prime}_{N}}}
(\vec{r}_{1},... \vec{r}_{N})
\psi_{\beta_{1},\beta_{2},...\beta_{N}}^{n_{1}^{\prime},n_{1}^{\prime},...n^{\prime}
_{N}}(\vec{r}_{1},... \vec{r}_{N}).
$$
 The indices $\beta_{1},...\beta_{N} $ are different for the closed shell states .
 If the wave function of the nucleon configuration is the Slater determinant, then
$$
\int d1...dN {\Psi^{\ast}}_{\beta_{1},...\beta_{N}}(\vec{r}_{1},...
\vec{r}_{N};m_{\nu{1}}^{\prime}...m_{\nu_{N}}^{\prime})
\Psi_{\beta_{1},...\beta_{N}}(\vec{r}_{1},...
\vec{r}_{N};m_{\nu{1}}^{\prime}...m_{\nu_{N}}^{\prime})=
  $$
$$
(1+ \sum_{{n_{1}^{\prime}\neq N,...n^{\prime}_{N}}}\int d\vec{r}_{1}...
d\vec{r}_{N}{\psi^{\ast}}_{\beta_{1},...\beta_{N}}^{n_{1}^{\prime},...n^{\prime}_{N}}
(\vec{r}_{1},...
\vec{r}_{N})\psi_{\beta_{1},...\beta_{N}}^{n_{1}^{\prime},...n^{\prime}_{N}}
(\vec{r}_{1},... \vec{r}_{N})).
$$
It is necessary to introduce a normalization factor
$$
  C=\sqrt{\frac{1}{1+ \sum_{n_{1}^{\prime}\neq N,...n^{\prime}_{N}} W_{n_{1}^{\prime}\neq
  N,...n^{\prime}_{N}}}}
 $$
 to obtain the full wave function $ \Psi_{\beta_{1},...\beta_{N}}(\vec{r}_{1},...
\vec{r}_{N};m_{\nu{1}}^{\prime}...m_{\nu_{N}}^{\prime}) $  normalized on 1.
 Here,
 $$
  W_{{n_{1}^{\prime}\neq N,...n^{\prime}_{N}}}=
  \int d\vec{r}_{1}...
d\vec{r}_{N}{\psi^{\ast}}_{\beta_{1},...\beta_{N}}^{n_{1}^{\prime}\neq
N,...n^{\prime}_{N}} (\vec{r}_{1},...
\vec{r}_{N})\psi_{\beta_{1},...\beta_{N}}^{n_{1}^{\prime}\neq N,...n^{\prime}_{N}}
(\vec{r}_{1},... \vec{r}_{N}).
 $$
The wave function of the space, spin, and isospin coordinates  of
the isobar configuration with quantum numbers $n_{1}^{\prime}\neq
N,...n^{\prime}_{N} $ is then
 $$
 {\psi^{\prime}}_{\beta_{1},...\beta_{N}}^{n_{1}^{\prime}\neq N,...n^{\prime}_{N}}
  (\vec{r}_{1},... \vec{r}_{N})=C \psi_{\beta_{1},...\beta_{N}}^{n_{1}^{\prime}
  \neq N,...n^{\prime}_{N}} (\vec{r}_{1},... \vec{r}_{N}), $$
  where $\psi_{\beta_{1},...
  \beta_{N}}^{n_{1}^{\prime}\neq N,...n^{\prime}_{N}} (\vec{r}_{1},... \vec{r}_{N})$
  is a  solution of the  Schr$\ddot{o}$dinger equation.

 It is sufficiently to deal with  a two-body wave function
 of the $\Delta$N system to find a momentum distribution of the delta- isobar in the nucleus.
 Let us consider the only one delta- isobar configuration. We write
 an antisymmetrization  operator as follows
$$
A=\frac{1}{\sqrt{N}}(1-\sum_{k=2}^{N}P_{1k})\frac{1}{\sqrt{N-1}}(1-\sum_{k=3}^{N}
P_{2k}) A_{3N}, $$ where $A_{3N} $ is the antisymmetrization
operator  of N-2 particles.  One may start from the two- body pair
wave function expansion $$ A_{3N}
 \psi_{\beta_{1},...\beta_{N}}^{\Delta(1),N(2),...N(N)} (\vec{r}_{1},... \vec{r}_{N})=
 \sum_{\kappa_{3},...\kappa_{N}} \psi^{\Delta (1)N(2)}_{\beta_{1},\beta_{2};\beta_{3}...
 \beta_{N},\kappa_{3},...\kappa_{N}} (\vec{r_{1}},\vec{r_{2}}) A_{3N} \psi_{\kappa_{3},...
 \kappa_{N}}^{N(3),...N(N)}
(\vec{r}_{3},... \vec{r}_{N}).
$$
Here,
 $A_{3N} \psi_{\kappa_{3},...\kappa_{N}}^{N(2),...N(N)}
  (\vec{r}_{3},... \vec{r}_{N})$
  forms a complete set of the  antisymmetric eigenfunctions of the N-2-
  particles system with energies
  $E_{\kappa_{3},...\kappa_{N}}$.
Substituting this expansion in the Schr$\ddot{o}$dinger equation , multiplying the
right- hand side  of the this equation on
 $A_{3N}
 \psi_{\alpha_{3},...\alpha_{N}}^{N(3),...N(N)} (\vec{r}_{3},... \vec{r}_{N})$,
   and using the condition of the orthogonality, one obtains
  the equation for the wave functions of the $\Delta N$ system
 $$
\sqrt{\frac{1}{N(N-1)}} \sum_{m_{\nu_{1}}...m_{\nu_{N}
 }; m_{\nu^{'}_{1}}...m_{\nu^{'}_{N}}}
 \phi_{\Delta(1),N(2)...N(N)}( m_{\nu_{1}}...m_{\nu_{N}})
 $$
 $$
(T(1)+T(2)+\Delta M_{1}- E_{\beta_{1},...\beta_{N}}+ E_{\alpha_{3},...\alpha_{N}}
)_{m_{\nu_{1}}...m_{\nu_{N}
 }; m_{\nu^{'}_{1}}...m_{\nu^{'}_{N}}}
 $$
 $$
 \phi_{\Delta(1),N(2)...N(N)}(m_{\nu{1}}^{\prime}...m_{\nu_{N}}^{\prime})
 \psi^{\Delta
(1),N(2)}_{\beta_{1},\beta_{2};\beta_{3}...\beta_{N},\alpha_{3},...\alpha_{N}}
(\vec{r_{1}}, \vec{r_{2}}) =
 $$
  $$
-\sum_{m_{\nu_{1}}...m_{\nu_{N}
 }; m_{\nu^{'}_{1}}...m_{\nu^{'}_{N}}}
 \phi_{\Delta(1),N(2)...N(N)}( m_{\nu_{1}}...m_{\nu_{N}})
(V_{12})_{m_{\nu_{1}}...m_{\nu_{N}
 }; m_{\nu^{'}_{1}}...m_{\nu^{'}_{N}}}
 \phi_{N(1),...N(N)}(m_{\nu{1}}^{\prime}...m_{\nu_{N}}^{\prime})
 $$
 $$
 \int d\vec{r}_{3},... d\vec{r}_{N} A_{3N}
 \psi_{\alpha_{3},...\alpha_{N}}^{N(3),...N(N)}
(\vec{r}_{3},... \vec{r}_{N})
 A_{N}
\cdot \psi_{\beta_{1},...\beta_{N}}^{N(1),...N(N)} (\vec{r}_{1},... \vec{r}_{N}).
 $$
 As the wave function of the $\Delta N$ system and
 the wave function  $\psi_{\beta_{1},...\beta_{N}}^{\Delta(1),N(2)...N(N)}
(\vec{r}_{1},... \vec{r}_{N}) $ are normalized to the same
constant, the solution of the  Schr$\ddot{o}$dinger equation for
the bound $\Delta N$ system must
 be multiplied by the above mentioned constant C.
 Solving this equation for the $ \Delta N$ system  in the shell model with the
 ls- coupling, one can obtain
  $$
  \psi^{\Delta
(1)N(2)}_{\beta_{1},\beta_{2};\beta_{3}...\beta_{N},\alpha_{3},...\alpha_{N}}
(\vec{r_{1}},\vec{r_{2}})=- G_{2}^{\Delta(1),N(2)}(E_{\alpha\beta})
$$
$$
  \sum_{m_{\nu_{1}}...m_{\nu_{N}
 }; m_{\nu^{'}_{1}}...m_{\nu^{'}_{N}}}
 \phi_{\Delta(1),N(2)...N(N)}( m_{\nu_{1}}...m_{\nu_{N}})
(V_{12})_{m_{\nu_{1}}...m_{\nu_{N}
 }; m_{\nu^{'}_{1}}...m_{\nu^{'}_{N}}}
 \phi_{N(1),...N(N)}(m_{\nu{1}}^{\prime}...m_{\nu_{N}}^{\prime})
 $$
 $$
  \sqrt{2}A_{12}\phi_{\beta_{1}}(\vec{r}_{1}) \phi_{\beta_{2}}(\vec{r}_{2})
    \delta_{\alpha_{3}\beta_{3}}...
    \delta_{\alpha_{N}\beta_{N}}.
  $$
  Here, $\phi_{\beta}(\beta=\{nlm_{l}sm_{s}tm_{t}\})$ are  single- particle
   states.
   This approach allows to use potentials, which depend from the
   relative coordinates of nucleons and isobars.
  The two- particle propagator $G_{2}^{\Delta(1),N(2)}(E_{\alpha\beta})$ is
  $$
  G_{2}^{\Delta(1),N(2)}(E_{\alpha\beta})=[\sum_{m_{\nu_{1}}...m_{\nu_{N}
 }; m_{\nu^{'}_{1}}...m_{\nu^{'}_{N}}}
 \phi_{\Delta(1),N(2)...N(N)}( m_{\nu_{1}}...m_{\nu_{N}})
 $$
 $$
(\frac{P^{2}_{1}}{2M_{1}}+\frac{P^{2}_{2}}{2M_{2}}+\Delta
M_{1}-E_{\beta_{1},...\beta_{N}}+ E_{\alpha_{3},...\alpha_{N}}
)_{m_{\nu_{1}}...m_{\nu_{N}
 }; m_{\nu^{'}_{1}}...m_{\nu^{'}_{N}}}
 \phi_{\Delta(1),N(2)...N(N)}(m_{\nu{1}}^{\prime}...m_{\nu_{N}}^{\prime})]^{-1}.
  $$
  The resulting  wave function of the $\Delta N$ system  in the closed shell
   nuclei with the ls- coupling is given in an explicit form in the work [11](s. Appendix 1.)

  Formulae given in work [11] can be used for the closed shell nuclei $^{4}He$,$^{16}O$.
  However, the nucleus $^{12}$C  has not
 the closed 1p shell in simple shell- model description  with the ls-coupling. To build a
 wave function of the isobar configuration with one delta in the case of
 the nucleus $^{12}$C, we used the shell model with the jj- coupling, in which the nucleus
 $^{12}$C has closed $s_{1/2}$ and  $p_{3/2}$ shells. As a radial parts of the one- particle
 wave functions for the case the jj- coupling and
the case the ls- coupling coincide for oscillator potential, the relationship
between these wave functions is defined by the angular parts of the wave functions.
 Result, we have
$$ \phi _{n_i^{^{\prime }}l_i^{^{\prime
}}j_i^{\prime}m_{j_i^{^{\prime }}}t_i^{^{\prime
}}m_{t_i}^{^{\prime }}}(\vec{r})= \sum
_{m_{l^{\prime}_{i}}m_{s_i^{^{\prime }}}}C_{ l_i^{^{\prime
}}m_{l_i^{^{\prime }}}s_i^{^{\prime }}m_{S_i^{^{\prime }}}}^{
j_i^{^{\prime }}m_i^{^{\prime }}} \phi _{n_i^{^{\prime
}}l_i^{^{\prime }}m_{l_i^{^{\prime }}}s_i^{^{\prime
}}m_{S_i}^{^{\prime }}t_i^{^{\prime }}m_{t_i^{^{\prime
}}}}(\vec{r}). $$ It allows to obtain a correlation between the
wave functions for the $\Delta N$ system in the case of the shell
model with the jj-and the ls-couplings $$ \Psi _{\alpha
_1^{^{\prime }}\alpha _2^{^{\prime }}}^{\Delta N}(1,2)= \sum
_{m_{l_1^{^{\prime }}}m_{l_2^{^{\prime }}}} \sum
_{m_{S_1^{^{\prime }}}m_{S_2^{^{\prime }}}}C_{ l_1^{^{\prime
}}m_{l_1^{^{\prime }}}s_1^{^{\prime }}m_{S_1^{^{\prime }}}}^{
j_1^{^{\prime }}m_1^{^{\prime }}} C_{ l_2^{^{\prime
}}m_{l_2^{^{\prime }}}s_2^{^{\prime }}m_{S_2^{^{\prime }}}}^{
j_2^{^{\prime }}m_2^{^{\prime }}} \Psi _{\beta _1^{^{\prime
}}\beta _2^{^{\prime }}}^{\Delta N}(1,2). $$
 Here, $ \alpha_{i}'=n_i^{^{\prime }}l_i^{^{\prime
}}j_i^{\prime}m_{{j_i^{^{\prime }}}}t_i^{^{\prime }}m_{{t_i}^{^{\prime }}}$, $\beta
_i^{^{\prime }}=n_i^{^{\prime }}l_i^{^{\prime }}m_{{l_i}^{^{\prime }}}s_i^{^{\prime
}}m_{{s_i}^{^{\prime }}}t_i^{^{\prime }}m_{{t_i}^{^{\prime }}}$.
  The wave functions of the $\Delta N$ system  for
 shell model with the jj- coupling is given in an explicit form  in Appendix 1.
\newpage
 \begin {center}
\textbf { $\Delta$- isobar momentum distribution in  nuclei}
\end {center}

By definition, the $\Delta $- isobar momentum distribution can be written through
the wave function of the $\Delta N$ system  as follows
 $$
\Delta _{\delta _{1}^{\prime }\delta _{2}^{\prime }}(\vec{k})=\int d {\hat{\vec{k}
}}d{\vec{p}}_{1}d{\vec{p}}_{2}{{\Psi }^{+}}_{\delta _{1}^{\prime } \delta
_{2}^{\prime }}^{\Delta N}({\vec{p}}_{1},{\vec{p}} _{2})\delta ({\vec{k}}-{
\vec{p}}_{1})\Psi _{\delta _{1}^{\prime }\delta _{2}^{\prime }}^{\Delta N}({
\vec{p}}_{1},{\vec{p}}_{2}).
$$
The indices $ \delta _ {i} ^ {\prime} $ (i = 1) are identified with the indices  $
\beta _ {i} ^ {\prime} $ in the case of the ls-coupling and the indices  $ \alpha _
{i} ^ {\prime} $ in the case of the jj-coupling. Here, $ \Psi _ {\delta _ {1} ^
{\prime} \delta _ {2} ^ {\prime}} ^ {\Delta N}
 ({\vec{p}} _ {1}, {\vec{p}} _ {2}) $ is the Fourier transform of the
  wave function of the  $\Delta N$  system

$$ \Psi ^ {\Delta N} _ {\delta _ 1 ^ {\prime}
 \delta _ 2 ^ {\prime
}} ({\vec{p}} _ 1, {\vec{p}} _ 2) = \frac {1} {(2\pi) ^ {3}} \int d {\vec{r} _ {1}}
d {\vec{r} _ {2}} \Psi ^ {\Delta N} _ {\delta _ 1 ^ {\prime} \delta _ 2 ^ {\prime}}
({\vec{r}} _ 1, {\vec{r}} _ 2) e ^ {-i {\vec{p}} _ {1} {\vec{r}} _ {1}} e ^ {-i
{\vec{p} } _ {2} {\vec{r}} _ {2}}.
$$
Using the wave functions  mentioned above, we have
$$\Psi _ {\delta _ 1 ^ {^ {\prime}} \delta _ 2 ^
{^ {\prime}}} ^ {\Delta N} ({\vec{p}} _ 1, {\vec{p}} _ 2) = \sum _ {\alpha} \Psi ^
 { \Delta N}_{ \alpha  {\delta _ 1 ^ {^ {\prime}} \delta _ 2 ^
 {^ {\prime}}}} ({\vec{p}}) \Phi _ {\alpha} ({\vec{P}}),
$$
where $ \alpha = N ^ {^ {\prime}} L ^ {^ {\prime}} M _ {L ^ {^ {\prime}}} $.
 Here, $ \Psi ^
 { \Delta N}_{ \alpha  {\delta _ 1 ^ {^ {\prime}} \delta _ 2 ^
 {^ {\prime}}}} ({\vec{p}})  $
 is the Fourier transform of the function of relative motion
 and $ \Phi _ {\alpha} ({\vec{P}}) $
  is the Fourier transform of the wave function of c.m. system

$$ \Psi ^ {\Delta N}  _ {N ^ {\prime} L ^ {\prime} M _ {L ^ {\prime}}{\delta _ {1} ^ {\prime} \delta _ {2}
 ^ {\prime}}} ({\vec{p}}) =
 \sum _ {\gamma ^ {\prime}, lsJM _ {J} TM _ {T}} \Psi ^ { \Delta N
  ; JM _ {J} TM _ {T} ls}
   _ {N ^ {\prime} L ^ {\prime} M _ {L ^ {\prime}}\delta _ {1} ^ {\prime}
    \delta _ {2} ^ {\prime} ; \gamma ^ {\prime}} (p) < \hat {\vec{p}}
    \mid (ls) JM _ {J} TM _ {T} >,
$$
     where
 $$
 \Psi ^ { \Delta N ; JM _ {J} TM _ {T} ls}
   _ {N ^ {\prime} L ^ {\prime} M _ {L ^ {\prime}}\delta _ {1} ^ {\prime}
    \delta _ {2} ^ {\prime} ; \gamma ^ {\prime}} (p)
  = F_{\delta_1^{^{\prime
\prime }}\delta_2^{^{\prime \prime}}s^{^{\prime }}l^{^{\prime
}}\Lambda ^{^{\prime }}}^{JM_JL^{^{\prime }}M_{L^{^{\prime }}}}
C_{ t_1^{^{\prime }}m_{t_1^{^{\prime }}}t_2^{^{\prime
}}m_{t_2^{^{\prime }}}}^{ Tm_{T}}a_{n^{^{\prime }}l^{^{\prime
}}N^{^{\prime }}L^{^{\prime }}}^{n_1^{^{\prime }}l_1^{^{\prime
}}n_2^{^{\prime }}l_2^{^{\prime }}\Lambda ^{^{\prime
}}}(1-(-)^{l^{^{\prime }}+s^{^{\prime }}+T})W_{ls,n^{\prime
}l^{\prime }s^{\prime }}^{\Delta NJTM_{J}M_{T}}(p), $$ $$ W _ {ls,
n ^ {^ {\prime}} l ^ {^ {\prime}} s ^ {^ {\prime}}} ^ {\Delta
NJTM_{J}M_{T}} (p) = \frac {1} {(2\pi) ^ {\frac {3} {2}}} \int drr
^ {2} 4\pi (-i) ^ {l} j _ {l} (pr) W _ {ls, n ^ {^ {\prime}} l ^
{^ {\prime}} s ^ {^ {\prime}}} ^ {\Delta NJTM_{J}M_{T}} (r),
 $$
$$
\Phi _ {N ^ {\prime} L ^ {\prime} M _ {L ^ {\prime}}} ({\vec{P}}) = R _ {N ^
{\prime}
 L ^ {\prime}} (P) < \hat {\vec{P}} \mid L ^ {\prime} M _ {L ^ {\prime}}
>,
$$
 and
$$ R _ {N ^ {\prime} L ^ {\prime}} (P) = \frac {1} {(2\pi) ^ {\frac {3} {2}}}
 \int drr ^ {2} 4\pi (-i) ^ {l} j _ {l} (Pr) R _ {N ^ {\prime} L ^ {\prime}} (r).
 $$
 The indices  $ \delta _ {i} ^ {\prime\prime} $ (i = 1) are identified with the
indices $\beta^{\prime\prime}_{i}=l^{\prime}_{i}m_{l^{\prime
}_{i}}s^{\prime }_{i}m_{s^{\prime }_{i}}$ in the case of the
ls-coupling and the indices $ \alpha _ {i} ^ {\prime\prime} $=
$l_i^{^{\prime }}j_i^{\prime}m_{j_i^{^{\prime }}}s_i^{^{\prime }}$
in the case of the jj-coupling. The momentum distribution
integrated over  angular variables is considered. Then, we can to
write the momentum distribution of the $\Delta$- isobar as follows
$$ \Delta_ {\delta _ {1} ^ {^ {\prime}} \delta _ {2} ^ {\prime}}
(k) =
 \sum _ {N ^ {^ {\prime}} L ^ {^ {\prime}} M _ {L ^ {\prime}};
 {\tilde N ^ {\prime}} {\tilde L} ^ {\prime} {\tilde M} _ {L ^ \prime}}
 \sum _ {\gamma ^ {\prime}, lsJTM _ {J} M _ {T};
 {\tilde {\gamma} ^ {\prime}}, {\tilde l} {\tilde s} {\tilde J} {\tilde T}
 {\tilde M} _ {J} {\tilde M} _ {T}} \int p ^ {2} dpP ^ {2} dP
 \Psi ^ { +\Delta N ; JM _ {J} TM _ {T} ls}
   _ {N ^ {\prime} L ^ {\prime} M _ {L ^ {\prime}}\delta _ {1} ^ {\prime}
    \delta _ {2} ^ {\prime} ; \gamma ^ {\prime}} (p)
  $$
  $$ \Psi ^ { \Delta N ; JM _ {J} TM _ {T} ls}
   _ {N ^ {\prime} L ^ {\prime} M _ {L ^ {\prime}}\delta _ {1} ^ {\prime}
    \delta _ {2} ^ {\prime} ; \gamma ^ {\prime}} (p)
    R _ {N ^ {\prime} L ^ {\prime} M _
    {L ^ {\prime}}} (P) R _ {{\tilde N} ^ {\prime} {\tilde L} ^ {\prime} {\tilde M} _
    {L ^ {\prime}}} (P)
 $$
$$
A _ {{\tilde {L}} ^ {\prime} {\tilde M} ^ {\prime}; ({\tilde l} {\tilde s}) {\tilde
J} {\tilde M} _ {J} {\tilde T} {\tilde M} _ {T}} ^ {L ^ {\prime} M _ {L ^ {\prime}};
(ls) JM _ {J} TM _ {T}} (k, p, P),
$$
where  $ A _ {{\tilde {L}} ^ {\prime} {\tilde M} ^ {\prime}; ({\tilde l} {\tilde s})
{\tilde J} {\tilde M} _ {J} {\tilde T} {\tilde M} _ {T}} ^ {L ^ {\prime} M _ {L ^
{\prime}}; (ls) JM _ {J} TM _ {T}} (k, p, P) $ is  the integral over the angular
variables

$$ A _ {{\tilde {L}} ^ {\prime} {\tilde M} ^ {\prime}; ({\tilde l}
{\tilde S}) {\tilde J} {\tilde M} _ {J} {\tilde T} {\tilde M} _
{T}} ^ {L ^ {\prime} M _ {L ^ {\prime}}; (ls) JM _ {J} TM _ {T}}
(k, p, P) = \int d {\hat {\vec{k}}} d {\hat {\vec{p}}} d {\hat
{\vec{P}}} < JM _ {J} (ls) TM _ {T} | {\hat {\vec{p}}} > < L ^ {^
{\prime}} M _ {L ^ {^ {\prime}}} | {\hat {\vec{P}}} >
 $$
$$ \delta (\vec{k}-\vec{p}-\frac {M_{1}} {M} {\vec{P}}) < {\hat
{\vec{p}}} \mid {\tilde L}
 ^ {^ {\prime}} {\hat M} _ {L ^ {^ {\prime}}} > < {\hat {\vec{p}}} | {\hat l} {\tilde s})
  {\tilde J} {\tilde M} _ {J} {\tilde T} {\tilde M} _ {T} >.
   $$
It is possible to show that
$$ A _ {{\tilde L} ^ {\prime} {\tilde M} ^ {\prime}; ({\tilde l} {\tilde s})
{\tilde J} {\tilde M} _ {J} {\tilde T} {\tilde M} _ {T}} ^ {L ^ {\prime} M _ {L ^
{\prime}}; (ls) JM _ {J} TM _ {T}} (k, p, P) = (-1) ^ {s + J + \tilde {l}
 + \tilde {J}} \frac {2} {\pi} \delta _ {T {\tilde T}} \delta _ {M _ {T} {\tilde M}
  _ {T}} \delta _ {{\tilde s} s} {\hat l} {\hat L ^ {\prime}} {\hat {J}} {\hat {\tilde J}}
  \sum _ {\lambda m _ {\lambda}} (-1) ^ {\lambda- m _ {\lambda}} {\hat \lambda} ^ {2}
$$
$$
C ^ {\tilde {L '} 0} _ {L'0\lambda 0} C ^ {\tilde {l} 0} _ {l0\lambda 0}
 \left \{
\begin {array} {c}

\lambda {\tilde J} J \\
S l {\tilde l}
\end {array}
\right \}
$$
$$
(-1) ^ {\tilde {L '} -M _ {\tilde {L '}}} (-1) ^ {\tilde {J} -M _ {\tilde {J}}}
 \left (\begin {array} {c}
\lambda {\tilde J} J \\
-m _ {\lambda} -M _ {\tilde {J}} M _ {J}
\end {array}
\right) \left (
\begin {array} {c}
L ' \lambda \tilde {L '} \\
M _ {L '} m _ {\lambda} -M _ {\tilde {L '}}
\end {array}
\right)
$$

$$ \int q ^ {2} dqj _ {0} (kq) j _ {\lambda} (pq) j _ {\lambda}
(\frac {M_{1}} {M} Pq). $$
Below, the momentum distribution is considered as sum
over all possible magnetic quantum numbers of the closed shells of the initial
nucleons
$$\Delta _ {\overline {\delta} _ {1} ^ {\prime} \overline {\delta}
_ {2} ^ {\prime}}
 (k) =\sum \Delta _ {\delta _ {1} ^ {\prime} \delta _ {2} ^ {\prime}} (k),
$$
 where $
\overline {\delta} _ {i} ^ {\prime}=\overline {\beta} _ {i} ^ {\prime} = n _ i ^ {^
{\prime}} l _ i ^ {^ {\prime}} s _ i ^ {\prime}
 t _ i ^ {^ {\prime}} $ for the case of the ls- coupling and $ \overline
 {\delta} _ {i} ^ {\prime}=\overline {\alpha} _ {i} ^ {\prime} = n _ i ^ {^ {\prime}}
  l _ i ^ {^ {\prime}} j _ i ^ {\prime}
 t _ i ^ {^ {\prime}} $ for the case of the jj- coupling.
 The sum over the quantum numbers $ m _ {l _ {1} '} m _ {l _ {2} '} m _ {s _ {1} '}
  m _ {s _ {2} '} m _ {t _ {1} '} m _ {t _ {1} '} $ for the case of the ls- coupling
   and $ m _ {j _ {1} '} m _ {j _ {2} '} m _ {t _ {1} '} m _ {t _ {1} '} $
  for the case of the jj- coupling is supposed.
 After using of the graphic methods of the summation  over the projections of
  the angular  momenta [13], we can write the
expression for the momentum distribution for the cases of the ls- and the jj-
couplings in the forms, which are given in Appendix 2.
\begin {center}
\textbf {The momentum distributions of $ \Delta $ - isobar for the
nuclei $ ^ 4He $ and $ ^ {16} O $}
 \end {center}

For the simplest nucleus with the closed shell $ ^ 4He $ $ n _ {1}
^ {\prime} = n _ {2} ^ {\prime} = l _ {1} ^ {\prime} = L _ {2} ^
{\prime} = 0 $. Therefore, we have $ 2n ^ {\prime} + 2N ^ {\prime}
+ l ^ {\prime}
 + L ^ {\prime} = 0 $. As all terms are more than zero, $ n ^ {\prime} = N ^ {\prime}
 = l ^ {\prime} = L ^ {\prime} = 0 $. From here, the following rules of the vector
 addition for the angular momenta must fulfill
  $ (\tilde JJ\lambda),
  (\lambda 00),(J0s ^ {\prime}), (s ^ {\prime} {\tilde J} {\tilde l} ^ {\prime})$
  (the rule of triangle).
 Similarly, it have place  $ (00\lambda) , (\lambda 00), (00\Lambda ^ {\prime}),
 \Lambda ^ {\prime} 00) $.
  At last,  we have $ (\tilde JJ\lambda), (\lambda l {\tilde l}), Jls),
  (s {\tilde J} {\tilde l}) $ from the  third 6j- symbol.
The transition potential gives the limitations
    l = 0,  $ (J l s), (s2s ^ {\prime}), l2s ^ {\prime}), (0Js ^ {\prime}), (02l),
   (s _ {1} = 3/2s _ {2} = 1/2s) $. From here, we have
$$
\lambda = 0, \Lambda ^ {\prime} = 0, {\tilde J} = J, l = {\tilde l}, J = s ^
{\prime}, l = 0,2, s (s ^ {\prime}, l),
$$
where $ s (s ^ {\prime}, 0) = J = s ^ {\prime} $, $ s (0,2) = 2 $ because of $ (J =
0sl) $ and
 $ s (1,2) = 1,2 $ because of $ (J = 1sl) $. The latter follows from  the
  9j- symbol properties.

 As result, we shall receive
$$ \Delta _ {\overline {\beta} _ {1} ^ {\prime} \overline {\beta}
_ {2} ^ {\prime}} = \sum _ {s ^ {\prime} = 0} ^ {1} \sum _ {J = s
^ {\prime}} \sum _ {T = 0} ^ {1} \sum _ {l = 0,2} \sum _ {s (s ^
{\prime}, l)} M ^ {slJ s lJT} _ {00s ^ {\prime}; { 00s ^
{\prime}}} (0) $$ $$\int q ^ {2} dqj _ {0} (kq) \int P ^ {2} dPj _
{0} (\frac {M _ {\Delta}} {M} Pq) R _ {00} (P) ^ {2} \int p ^ {2}
dp j _ {0} (pq) {\mid W _ {ls, 00s ^ {\prime}} ^ {\Delta NJT} (p)
\mid} ^ {2}, $$ where $$ M ^ {slJ s lJT} _ {00s ^ {\prime}; {00s ^
{\prime}}} (0) = (-1) ^ {s + l} (1 - (-1) ^ {s ^ {\prime + T}}) ^
{2} \frac {2} {\pi} \hat {l} \hat {J} ^ {4} \hat {T} ^ {2} N ^
{slJlJ} _ {s ^ {\prime}
 000} (0)
$$
and
 $$
N^{slJlJ}_{s^{\prime}000}(0)=\{ JJ 0\}\{000\}\left\{
\begin{array}{c}
 JJ0\\
 00s^{\prime}
\end{array}
\right\} \left\{
\begin{array}{c}
 000\\
 000
\end{array}
\right\}\left\{
\begin{array}{c}
 JJ0\\
 lls
\end{array}
\right\}. $$ Using  properties of the 3j- and 6j- symbols, summing
over l
 and $ s', T $,  we shall receive
$$ \Delta _ {\overline {\beta} _ {1} ^ {\prime} \overline {\beta}
_ {2} ^ {\prime}} = \frac {24} {\pi} \int q ^ {2} dqj _ {0} (kq)
\int P ^ {2} dPj _ {0} (\frac {M _ {\Delta}}
 {M} Pq) R _ {00} (P) ^ {2} \int p ^ {2} dp j _ {0} (pq)
$$
$$
( ^ 1S _ {0} + ^ 5D _ {0} - ^ 3S _ {1} - ^ 3D _ {1} - ^ 5D _ {1}),
$$
where
$$
^ 1S _ {0} = {\mid W _ {00,000} ^ {\Delta N01} (p) \mid} ^ {2}, ^ 5D _ {0} = {\mid W
_ {22,000} ^ {\Delta N01} (p) \mid} ^ {2}
$$
$$
^ 3S _ {1} = {\mid W _ {01,001} ^ {\Delta N10} (p) \mid} ^ {2}, ^ 3D _ {1} = {\mid W
_ {21,001} ^ {\Delta N10} (p) \mid} ^ {2}, ^ 5D _ {1} = {\mid W _ {22,001} ^ {\Delta
N10}
 (p) \mid} ^ {2}).
$$
 It is possible to perform the q' integration of the last expression. Besides, it is
possible to show, that the matrix element of the transition
potential is not zero only for the value of $ ^ 5D _ {0} $. The
values $ ^ 3S _ {1} , ^ 3D _ {1}, ^ 5D _ {1}$ include the matrix
element of the transition  potential with T = 0, and so, they are
equal to zero because of (03/21/2). The value $ ^ 1S _ {0} $ is
equal to zero because of the rule for the vector addition
(03/21/2) for the spin angular momentum . Therefore, the momentum
distribution of the $\Delta$- isobar in the nucleus $ ^ 4He $  is
$$ \Delta _ {\overline {\beta} _ {1} ^ {\prime} \overline {\beta}
_ {2} ^ {\prime}} = \frac {24} {(2\pi) ^ {4}} \frac {1} {\alpha}
\sqrt {\frac {\pi} {2\nu}} \int p dp \frac {1} {k} (exp (-\frac
{(k-p) ^ {2}} {2\alpha ^ {2} \nu}) exp (-\frac {(k + p) ^ {2}}
{2\alpha ^ {2} \nu})) \widetilde {\mid W _ {22,000} ^ {\Delta N01}
(p) \mid ^ {2}}. $$ Here, $$ \widetilde{W_{22,000}^{\Delta
N01}(p)}=16\mu_{\Delta N}B_{\Delta N}\{\int_{0}^{\infty}
r^{2}drj_{2}(pr)k_{2}(B_{\Delta
N}r)\int_{0}^{r}{r^{\prime}}^{2}dr^{\prime}i_{2}(B_{\Delta
N}r^{\prime}V_{22,00}^{01}(r^{\prime})R_{00}(r^{\prime})+ $$ $$
\int_{0}^{\infty} r^{2}drj_{2}(pr)i_{2}(B_{\Delta
N}r)\int_{r}^{\infty}{r^{\prime}}^{2}dr^{\prime}k_{2}(B_{\Delta
N}r^{\prime})V_{22,01}^{01}(r^{\prime})R_{00}(r^{\prime})\}. $$
The numerical calculations of the momentum distribution of the
$\Delta$- isobar for the nucleus $ ^ {4} He $  are shown on Fig. 1
( curve  a). The probability for finding one $\Delta$- isobar per
nucleon for the nucleus $ ^ {4} He $ is given in Table 2.

 For the nucleus $ ^ {16} O $ the following sets of the main quantum numbers
and the angular momentums are possible. Because of a centrifugal
barrier two nucleons have the maximal probability to turn into  a
pair $ \Delta N $ only in the state with $ l ^ {\prime} $ = 0.
Therefore, the sum  over the initial quantum numbers are
simplified greatly. The possible initial quantum number are given
in Table 1.

\hspace{-0.3cm}{\begin{tabular}{|c|c|c|c|c|c|c|c|c|}\hline
$n_{1}^{\prime}=0$&$l_{1}^{\prime}=0$&$n_{2}^{\prime}=0$&$l_{2}^{\prime}=0$&$n^{\prime}=0$
&$N^{\prime}=0$&$l^{\prime}=0$&$L^{\prime}=0$&$\Lambda^{\prime}=0$\\
\hline
$n_{1}^{\prime}=0$&$l_{1}^{\prime}=1$&$n_{2}^{\prime}=0$&$l_{2}^{\prime}=0$&$n^{\prime}=0$
&$N^{\prime}=0$&$l^{\prime}=0$&$L^{\prime}=1$&$\Lambda^{\prime}=1$\\
\hline
$n_{1}^{\prime}=0$&$l_{1}^{\prime}=0$&$n_{2}^{\prime}=0$&$l_{2}^{\prime}=1$&$n^{\prime}=0$
&$N^{\prime}=0$&$l^{\prime}=0$&$L^{\prime}=1$&$\Lambda^{\prime}=1$\\
\hline
$n_{1}^{\prime}=0$&$l_{1}^{\prime}=1$&$n_{2}^{\prime}=0$&$l_{2}^{\prime}=1$&$n^{\prime}=0$
&$N^{\prime}=0$&$l^{\prime}=0$&$L^{\prime}=2$&$\Lambda^{\prime}=2$\\
\hline&&&&$n^{\prime}=1$
&$N^{\prime}=0$&$l^{\prime}=0$&$L^{\prime}=0$&$\Lambda^{\prime}=0$\\
\hline &&&&$n^{\prime}=0$
&$N^{\prime}=1$&$l^{\prime}=0$&$L^{\prime}=0$&$\Lambda^{\prime}=0$\\
\hline
\end{tabular} }\\\\
\hspace {6cm} Table 1. The initial quantum numbers for
 the nucleus $ ^ {16} O $.

If  $ l ^ {\prime} $ = 0, the rule of the vector addition of the  angular momentum $
(\lambda l ^ {\prime} \tilde {l ^ {\prime}}) $ gives $\lambda $=0.
 Then, we have
$$
M ^ {slJ \tilde s\tilde l {\tilde J} T} _ {N ^ {\prime} L ^ {\prime} \gamma ^
{\prime}; {\tilde N} ^ {\prime} {\tilde L} ^ {\prime} {\tilde\gamma ^ {\prime}}} (0)
= (-1) ^
 {s + \tilde l + {\tilde L} ^ {\prime} + J
-s ^ {\prime} -\Lambda of ^ {\prime}} ( 1 - (-1) ^ {s ^ \prime + T}) (1 - (-1) ^
{{\tilde s} ^ \prime + T}) \hat {T} ^ {2}
$$
$$
\frac {2} {\pi} \delta _ {s ^ {\prime} {\tilde s} ^ {\prime}} \delta _ {\Lambda ^
{\prime} {\tilde \Lambda} ^ {\prime}} \delta _ {s {\tilde s}} \hat {l} \hat {J} ^
{2} \hat
 {\tilde J} ^ {2} \hat {\Lambda ^ {\prime}} ^ {2} \hat {{L ^ {\prime}}} \hat {\lambda}
  ^ {2}
  C ^ {{\tilde L} ^ {\prime} 0} _ {L ^ {\prime} 0 \lambda 0} C ^ {00} _ {00 00}
$$
$$
a _ {n ^ {\prime} l ^ {\prime} N ^ {\prime} L ^ {\prime}} ^ {n _ 1 ^ {\prime} l _ 1
^ {\prime} n _ 2 ^ {\prime } L _ 2 ^ {\prime} \Lambda ^ {\prime}} a _ {{\tilde n} ^
{\prime} {\tilde l} ^ {\prime}
 {\tilde N} ^ {\prime} {\tilde L} ^ {\prime}} ^ {n _ 1 ^ {\prime} l _ 1 ^ {\prime}
  n _ 2 ^ {\prime} l _ 2 ^ {\prime} {\tilde \Lambda} ^ {\prime}} N ^ {slJ {\tilde l}
   {\tilde
J}} _ {s ^ {\prime} 0L ^ {\prime} \Lambda ^ {\prime} {0} \tilde {\Lambda} ^
{\prime}} (0),
$$
where
$$
N^{slJ{\tilde l}{\tilde J}}_{s^{\prime}0L^{\prime}\Lambda^{\prime}{0}\tilde {L}
^{\prime }\tilde {\Lambda} ^{\prime }}(0) =\{{\tilde J}J 0\}\{{\tilde L}^{\prime
}L^{\prime}0\left\{
\begin{array}{c}
{\tilde J} J0 \\
 00s^{\prime}
\end{array}
\right\} \left\{
\begin{array}{c}
{\tilde L}^{\prime}L^{\prime} 0  \\
00\Lambda^{\prime}
\end{array}
\right\}\left\{
\begin{array}{c}
{\tilde J} J0 \\
 l {\tilde l}s
\end{array}
\right\}
$$
and $ \gamma ^ {\prime} = s ^ {'} \Lambda ^ {'} n ^ {'} l ^ {'} $. Using properties
the 3j-and 6j- symbols, we  receive
$$
N ^ {slJ {\tilde l} {\tilde J}} _ {s ^ {\prime} 0L ^ {\prime} \Lambda ^ {\prime} {0}
\tilde {L} ^ {\prime} \tilde {\Lambda} ^ {\prime}} (0) = (-1) ^ {s + \tilde l +
\tilde L ^ {\prime}
 + 2\tilde J + s ^ {\prime} + \Lambda ^ {\prime}} \delta _ {J\tilde {J}} \delta _
 {L ^ {\prime} {\tilde L} ^ {\prime}} \delta _ {l\tilde {l}} \frac {1} {\hat {\tilde J} ^ {2}
 \hat {l} \hat {\tilde {L ^ {\prime}}}}.
$$
 Hence (s. Appendix 2),
  $$ M ^ {slJ \tilde s\tilde l {\tilde J}
T} _ {N ^ {\prime} L ^ {\prime} \gamma ^ {\prime}; {\tilde N} ^
{\prime} {\tilde L} ^ {\prime} {\tilde\gamma ^ {\prime}}} (0) =
(-1) ^ { J} (1 - (-1) ^ {s ^ \prime + T}) ^ {2} $$ $$ \frac {2}
{\pi} \delta _ {s ^ {\prime} {\tilde s} ^ {\prime}} \delta _
{\Lambda ^
 {\prime} {\tilde \Lambda} ^ {\prime}} \delta _ {L ^ {\prime} {\tilde
L} ^ {\prime}} \delta _ {s {\tilde s}} \delta _ {l\tilde L} \delta _ {J\tilde J}
\hat {T} ^ {2} \hat {J} ^ {2} \hat {\Lambda ^ {\prime}} ^ {2}
$$
$$
a _ {n ^ {\prime} l ^ {\prime} N ^ {\prime} L ^ {\prime}} ^ {n _ 1 ^ {\prime} l _ 1
^ {\prime} n _ 2 ^ {\prime } L _ 2 ^ {\prime} \Lambda ^ {\prime}} a _ {{\tilde n} ^
{\prime} {\tilde l} ^ {\prime} {\tilde N} ^ {\prime} {\tilde L} ^ {\prime}} ^ {n _ 1
^ {\prime} l _ 1 ^ {\prime} n _ 2 ^ {\prime} l _ 2 ^ {\prime} {\tilde \Lambda} ^
{\prime}},
$$
where $ \gamma ^ {\prime} = s ^ {'} \Lambda ^ {'} n ^ {'} (l ^ {'} = 0) $. It gives
the following expression for the momentum distribution of the $ \Delta $- isobar in
the nucleus $ ^ {16} O $
$$
\Delta _ {\overline {\beta} _ {1} ^ {\prime} \overline {\beta} _ {2} ^ {\prime}} =
\sum _
 {N ^ {'} {\tilde N} ^ {\prime} L ^ {'} {\tilde L} ^ {\prime} n ^ {\prime} \tilde
 {n ^ {\prime}} s ^ {\prime} \Lambda ^ {\prime} {\tilde \Lambda} ^ {\prime}}
 \sum _ {lsJT} (-1) ^ {-J + L ^ {\prime} - {\tilde L} ^ {\prime}}
 (1 - (-1) ^ {s ^ \prime + T}) ^ {2} \frac {2} {\pi} \hat {T} ^ {2}
 \hat {J} ^ {2} \hat {\Lambda ^ {\prime}} ^ {2} a _ {n ^ {\prime} l ^
 {\prime} N ^ {\prime} L ^ {\prime}} ^ {n _ 1 ^ {\prime} l _ 1 ^
 {\prime} n _ 2 ^ {\prime
} L _ 2 ^ {\prime} \Lambda ^ {\prime}} a _ {{\tilde n} ^ {\prime} {\tilde l} ^
{\prime} {\tilde N} ^ {\prime} {\tilde L} ^ {\prime}} ^ {n _ 1 ^ {\prime} l _ 1 ^
{\prime} n _ 2 ^ {\prime } L _ 2 ^ {\prime} {\tilde \Lambda} ^ {\prime}} \delta _ {L
^ {\prime} {\tilde L} ^ {\prime}} \delta _ {\Lambda ^ {\prime} {\tilde \Lambda} ^
{\prime}}
$$
$$\int q ^ {2} dqj _ {0} (kq) \int P ^ {2} dPj _ {0} (\frac {M _ {\Delta}} {M} Pq)
 R _ {N ^ {\prime} L ^ {\prime}} R _ {{\tilde N} ^ {\prime} {\tilde L} ^ {\prime}}
  (P)
  \int p ^ {2} dp j _ {0} (pq) W _ {ls, n ^ {\prime} l ^ {\prime} s ^ {\prime}}
  ^ {\Delta NJT} (p) W _ {{l} s, {\tilde n} ^ {\prime} {l} ^ {\prime} {s}
   ^ {\prime}} ^ {\Delta N {J} T} (p),
$$ where $ l ^ {\prime} = {\tilde l} ^ {\prime} = 0 $. The sum
over the initial quantum numbers is carried out according to the
Table 1 at the fixed $ l _ {i} ' $. Depending on a shell, where
the two initial nucleons turn into  a pair $ \Delta N $, there are
three different combinations of the values $ l _ {1} ' $, $ l _
{2} ' $.
 The sum momentum distribution of the $ \Delta$- isobar  is
$$
\Delta (k) = \Delta _ {01\frac {1} {2} \frac {1} {2}; 01\frac {1} {2}
 \frac {1} {2}} (k) + \Delta _ {01\frac {1} {2} \frac {1} {2}; 00\frac {1
 } {2} \frac {1} {2}} (k) + \Delta _ {00\frac {1} {2} \frac {1} {2}; 01\frac {1}
  {2} \frac {1} {2}} (k) + \Delta _ {00\frac {1} {2} \frac {1} {2}; 00\frac {1} {2}
   \frac {1} {2}} (k).
$$
The sum over $ s ^ {\prime} $, l, s, J, T is carried out as well as
 in the case of the nucleus $ ^ 4He $. The identical the limitations follow because of the
 same transition potential. In this case  $ l = 0,2, J = s ^ {\prime} $,
 $ s (s ^ {\prime}, 0) =  s ^ {\prime} $, $ s (0,2) = 2 $ because of $ (J = 0sl) $,
 $ s (1,2) = 1,2 $ because of $ (J = 1sl) $. As in the previous case,
  because of the rules of the vector addition (03/21/2)for T and (03/21/2)for the spin
  angular momentum, the transitions with T = 0  and the transitions with
  $ s' = 0, l = 0, s = 0, J = 0, T = 1 $ are absent. As result, the  momentum
   distribution of the $ \Delta$- isobar in the nucleus $ ^ {16} $ O is
 $$
\Delta (k) = \frac {24} {\pi} \int q ^ {2} dqj _ {0} (kq) \int P ^ {2} dPj _ {0}
(\frac {M _ {\Delta}} {M} Pq) \int p ^ {2} dpj _ {0} (pq) [
$$
$$
( R _ {00} (P) ^ {2}) \mid W _ {22,000} ^ {\Delta N01} (p) \mid ^ {2} +
$$
$$
+ 3R _ {01} (P) ^ {2} \mid W _ {22,000} ^ {\Delta N01} (p) \mid ^ {2} +
$$
$$
( \frac {5} {2} R _ {02} (P) ^ {2} + \frac {1} {2} R _ {10 (P) ^ {2}}) {\mid W _
{22,000} ^ {\Delta N01} (p) \mid} ^ {2} -R _ {10} (P) R _ {00} (P) W _ {22,000} ^
{\Delta N01} (p) W _ {22,10 0} ^ {\Delta N01} (p)
 + \frac {1} {2} R _ {00} (P) ^ {2} \mid W _ {22,100} ^ {\Delta N01} (p) \mid ^ {2}].
$$

The Fourier transform of the radial parts are determined by  formula
$$
R _ {nl} (p) = \frac {4\pi} {(2\pi) ^ 3/2} (-i) ^ {l} \int r ^ {2} drj _ {l} (pr) R
_ {nl} (r) .
$$
After an analytical calculation of the available integrals, we shall receive the
following expression for the  momentum distribution
 of the $ \Delta$- isobar in the nucleus $ ^ {16}O $
$$ \Delta(k)= \frac{24}{(2\pi)^{4}}\frac{1}{\alpha}
\sqrt{\frac{\pi}{2\nu}}\frac{1}{k}\int pdp $$ $$
(exp(-\frac{(k-p)^{2}}{2\alpha^{2}\nu})f_{1}(k-p)-
exp(-\frac{(k+p)^{2}}{2\alpha^{2}\nu})f_{1}(k+p) {\mid\widetilde{
W_{22,000}^{\Delta N01}(p)}\mid}^{2} $$ $$
+\sqrt{6}(exp(-\frac{(k-p)^{2}}{2\alpha^{2}\nu})f_{2}(k-p)-
exp(-\frac{(k+p)^{2}}{2\alpha^{2}\nu}f_{2}(k+p)
))\widetilde{W_{22,000}^{\Delta N01}(p)}\widetilde{W_{ 22,10
0}^{\Delta N01}(p)} $$ $$
+(exp(-\frac{(k-p)^{2}}{2\alpha^{2}\nu})f_{3}(k-p)-
exp(-\frac{(k+p)^{2}}{2\alpha^{2}\nu}f_{3}(k+p)  )\mid\widetilde{
W_{22,100}^{\Delta N01}(p)}\mid^{2}),
 $$ where
   $$ f_{1}(x)=6
+\frac{1}{4}\frac{x^{2}}{\alpha^{2}\nu}+
\frac{1}{4}\frac{x^{4}}{\alpha^{4}\nu^{2}} $$ $$
f_{2}(x)=\frac{1}{6}(1-\frac{x^{2}}{\alpha^{2}\nu}) $$ $$
f_{3}(x)=\frac{1}{2}. $$ Here,
$\alpha=\frac{M_{\Delta}}{M_{\Delta}+ M_{N}}$

 Using the formula written above, we have
obtained the momentum distribution of the $\Delta$- isobar for the nucleus $ ^ {16}
O $. Results are shown on Fig. 1 ( curve b). The probability for finding
 of the $\Delta$- isobar  per nucleon in the nucleus $ ^ {16} O $ is
given in Table 2.
\begin {center}
\textbf {Momentum distribution of $ \Delta $ - isobar for the
nucleus $ ^ {12}C$}
\end {center}
Let us consider the momentum distribution of the $ \Delta $ - isobar in the ground
state of the nucleus $ ^ {12} $ C. This nucleus in the ground state have two closed
$ s _ {1/2} $ and $ l _ {3/2} $ shells. The sum over the quantum number for the
momentum distribution of the $ \Delta $ - isobar for the nucleus $ ^ {12} C $ (s.
the formulae of Appendix 2) is defined by the rules of the vector addition for the
angular momentums. Below, we shall consider the case $ l ' = 0 $.  By definition, we
have $ s _ {1} ' = \frac {1} {2}, s _ {2} ' = \frac {1} {2}, s _ {1} = \frac {3}
{2}, s _ {2} = \frac {1} {2} $. From the form of the transition potential it follows
l = 0 or l = 2. For these two cases $ s' = 0,1 $. From the 6j- symbols in the
isotopic space it follows T = 1. Then, because of the factor $ (1 - (-1) ^ {l ^
\prime + s ^ \prime + T}) $, the selection rule for spin  follows $ s' =\tilde {s'}
= 0 $. It is very strong limitation. We have  $ J = 0 $ and $ \lambda = 0 $. In
addition, $ l = \tilde {l} $ and $ s = l $. It follows  $ x = L ' = \tilde {L '} $ =
$  \Lambda ' = \tilde {\Lambda '} $ also. Because of the first part of the
transition potential the value l = 0. However, because of (03/21/2) the contribution
of this item disappears. Therefore, only $ l = s = 2 $ remains. Thus, the expression
for the momentum distribution becomes considerably simpler
$$
\Delta _ {\overline
{\alpha} _ {1} ^ {\prime} \overline {\alpha} _ {2} ^ {\prime}} (k) = \sum _ {N ^
\prime n ^ \prime {\tilde N} {\tilde n} ^ \prime} \sum _ {L ^ \prime \Lambda ^
\prime} M _ {N ^ \prime L ^ \prime n ^ \prime l ^ \prime = 0 s ^ \prime = 0 \Lambda
^ \prime; {\tilde N} ^ \prime {L} ^ \prime {\tilde n} ^ \prime {l} ^ \prime = 0 {s}
^ \prime = 0 {\Lambda} ^ \prime} ^ {\lambda = 0 l = s = 2J = 0\tilde l = 2\tilde J =
0T = 1} (\overline {\alpha} _ {1} ^ {\prime} \overline {\alpha} _ {2} ^ {\prime})
\int q ^ 2dq j _ {0} (kq)
 $$
$$
 \int P ^ 2dP j _ {\lambda} (\frac {M_{\Delta}} {M} Pq) R _ {N ^ {\prime} L ^ {\prime}}
(P) R _ {{\tilde N} ^ {\prime} {L} ^ {\prime}} (P) \int p ^ 2 dp j
_ {\lambda} (pq) W _ {l = 2s = 2n ^ {\prime} l ^ {\prime} = 0s ^
{\prime} = 0} ^ {\Delta NJ = 0T = 1} (p) W _ {{\tilde l} = 2s = 2,
{\tilde n} ^ {\prime} {\tilde l} ^ {\prime} = 2 {\tilde s} ^
{\prime} = 2} ^ {\Delta N {\tilde J} = 0T = 1} (p).
$$
 Let us  designate the momentum distributions, when both holes are on the
$p_{\frac{3}{2}}$ - shells, as follows
  $$ \Delta _ {p _ {\frac {3}
{2}} p _ {\frac {3} {2}}} = \Delta _ {\overline {\alpha} _ {1} ^ {\prime} \overline
{\alpha} _ {2} ^ {\prime}} (k),
 $$
 where we have $ \overline {\alpha} _ {1} ^
{\prime} $ $\equiv $ ($ n _ 1 ^ {\prime} = 0 l _ 1 ^ {\prime} = 1 j _ 1 ^ {\prime} =
3/2 t _ 1 ^ {\prime} = 1/2 $) $ \overline {\alpha} _ {2} ^ {\prime} $$\equiv $ ($ n
_ 2 ^ {\prime} = 0 l _ 2 ^ {\prime} = 1 j _ 2 ^ {\prime} = 3/2 t _ 2 ^ {\prime} =
1/2 $).
 Similarly, we shall designate the
momentum distribution, when one hole is on the $ s _ {\frac {1}
{2}} $, and other hole is on the $ p _ {\frac {3} {2}} $ - shell,
as follows
 $$ \Delta _ {p _ {\frac {1} {2}} p _ {\frac {3}
{2}}}=\Delta _ {\overline {\alpha} _ {1} ^ {\prime} \overline
{\alpha} _ {2} ^ {\prime}} (k),
 $$
 where $ \overline {\alpha} _
{1} ^ {\prime} $$\equiv $ ($ n _ 1 ^ {\prime} = 0 l _ 1 ^ {\prime}
= 0 j _ 1 ^ {\prime} = 1/2 t _ 1 ^ {\prime} = 1/2 $) $ \overline
{\alpha} _ {2} ^ {\prime} $$\equiv $ ($ n _ 2 ^ {\prime} = 0 l _ 2
^ {\prime} = 1 j _ 2 ^ {\prime} = 3/2 t _ 2 ^ {\prime} = 1/2 $).
 At last, we shall designate the momentum
distribution, when both holes are on the $ s _ {\frac {1} {2}} $ -
shells, as follows
 $$ \Delta _ {s _ {\frac {1} {2}} s _ {\frac {1}
{2}}} = \Delta _ {\overline {\alpha} _ {1} ^ {\prime} \overline
{\alpha} _ {2} ^ {\prime}} (k), $$ where $ \overline {\alpha} _
{1} ^ {\prime} $$\equiv $ ($ n _ 1 ^ {\prime} = 0 l _ 1 ^ {\prime}
= 0 j _ 1 ^ {\prime} = 1/2 t _ 1 ^ {\prime} = 1/2 $) $ \overline
{\alpha} _ {2} ^ {\prime} $$\equiv $ ($ n _ 2 ^ {\prime} = 0 l _ 2
^ {\prime} = 0 j _ 2 ^ {\prime} = 1/2 t _ 2 ^ {\prime} = 1/2 $).
Then, the momentum distribution of the $ \Delta$- isobars for  the
nucleus $ ^ {12} $ C is a sum of the momentum distributions for
three various cases
 $$ \Delta(k)=  \Delta _ {p _ {\frac {3} {2}} p
_ {\frac {3} {2}}}+  \Delta _ {p _ {\frac {3} {2}} p _ {\frac {2}
{3}}}+\Delta _ {p _ {\frac {2} {3}} p _ {\frac {3} {2}}}+ \Delta _
{s _ {\frac {1} {2}} s _ {\frac {1} {2}}}. $$
 Using the results of
the previous sections for a calculation of the integrals and the
values $ W _ {22,000} ^ {\Delta N01} (p)$ and  $W _ {22,100} ^
{\Delta N01} (p) $ with appropriate replacement of parameters of
oscillator model for $ ^ {16} O $ on parameters for $ ^ {12} C $,
we have $$
\Delta(k)=\frac{16}{(2\pi)^{4}}\frac{1}{2\alpha}\sqrt{\frac{\pi}{2\nu}}
\frac{1}{k} \int
pdp\{[exp(-\frac{(k-p)^{2}}{2\alpha^{2}\nu})f_{1}(k-p)-
exp(-\frac{(k+p)^{2}}{2\alpha^{2}\nu})f_{1}(k+p)]\tilde{{W_{22,000}^{\Delta
N01}(p)}^{2}}+ $$
 $$
 [exp(-\frac{(k-p)^{2}}{2\alpha^{2}\nu})f_{2}(k-p)-
exp(-\frac{(k+p)^{2}}{2\alpha^{2}\nu})f_{2}(k+p)]\tilde{W_{22,000}^{\Delta N01}(p)}
\tilde{W_{22,100}^{\Delta N01}(p)}+
 $$
 $$
 [exp(-\frac{(k-p)^{2}}{2\alpha^{2}\nu})f_{3}(k-p)-
exp(-\frac{(k+p)^{2}}{2\alpha^{2}\nu})f_{3}(k+p)]\tilde{{W_{22,100}^{\Delta
N01}(p)}^{2}}.
 $$
Here,
$$
f _ {1}(x)= \frac {11} {2} + 2 (2 + \frac {x ^ {2}} {\alpha ^ {2} \nu})+\frac {1}
{3} (2-\frac {x ^ {2}} {\alpha ^ {2} \nu}) ^ {2}
$$
$$
f _ {2}(x)= \frac {2} {\sqrt {6}} (1-\frac {x ^ {2}} {\alpha ^ {2} \nu})
$$
$$
f _ {3}(x)= 1.
$$
The numerical calculations of the momentum distribution of the $\Delta$- isobar for
the nucleus $ ^ {12} C $ are shown on Fig. 1 ( curve  c). The probability for
finding of the $\Delta$- isobar  per nucleon in the nucleus $ ^ {12} C $ is given in
Table 2.

 \hspace{-0.3cm} {

 \hspace{3cm}{\begin{tabular}{|c|c|c|c|c|c|}
 \hline
 Nucleus&$P_{1\Delta}$&C&$N$&$P_{total}$
&$P_{m}$\\
 \hline
$^{4}He$&2.73&0.8907&0.1227&0.1094
&0.1081\\
\hline $^{12}C$&2.19&0.8582&0.3574&0.2633
&0.2604\\
\hline $^{16}O$&2.17&0.8074&0.5338&0.3401
&0.3448\\
\hline
\end{tabular} }\\\\
 Table 2.
 The probability for finding
of the $\Delta$- isobar in the nucleus  per nucleon $ P_{1\Delta}$ is given in $\%$;
 C is the constant of the normalization;
N is the norm of the wave function of the $\Delta$- isobar; $
P_{t}$ is the full probability for finding of the $\Delta$- isobar
in the nucleus.
 $P_{m}$ is the result of the k
integration of the value $k^{2}\Delta(k)$.  }

          \begin{center}
  \textbf{Conclusions }
            \end{center}
The wave function of the $\Delta$- isobar configuration in closed shell nuclei  was
obtained in [11] in the harmonic oscillator model with the ls- coupling. The
transition potential with $\pi$- and $ \rho$- exchange was used. We use this result
for building of the wave function of the $\Delta$- isobar configuration in the case
of the harmonic oscillator model with the jj- coupling. The probability for finding
of the $\Delta$- isobar per nucleon in the nucleus and the momentum distribution of
the $\Delta$- isobar were calculated for the closed shell nuclei $^4 He$ and
$^{16}O$ in the case of the harmonic oscillator model with the ls- coupling and the
closed shell nucleus $^{12}C$ in the case of the harmonic oscillator model with the
jj- coupling. This results can be used for a interpretation of the experimental data
in reactions on nuclei in the new experiments connected with of the $ \Delta$
knock-out from the nuclei target by $\pi$- meson and photon beams. In particular,
the received results can be used for the interpretation of the experimental data in
the reaction $^{12}C(\gamma,\pi^{+}p)$ in the framework of the assumption that
formation of the $\pi^{+}p$ pairs may be interpreted as a $
\gamma\Delta^{++}\rightarrow\pi^{+}p$ process, which takes place on a $ \Delta^{++}$
preexisting in the target nucleus.

 \textbf{Appendix 1.} \\
a) The wave function of the $ \Delta $ N system obtained in  [11]
in the harmonic oscillator shell- model with the ls- coupling can
be presented as follows
  $$
  \Psi^{\Delta N}_{\beta_{1}^{'}\beta_{2}^{'}}( \vec{r_{1}},\vec{r_{2}})=
  \sum_{N^{'}L^{'}M_{L}^{'}} \Psi^{\Delta N}_{\beta_{1}^{'}\beta_{2}^{'}
  N^{'}L^{'}M_{L}^{'}}
 ({\vec{r}})\Phi_{N^{'}L^{'}M_{L}^{'}} ({\vec{R}}).
  $$
  The part of the wave function dependent on $ \vec {r} $ can be written as
   follows
 $$
 \Psi^{\Delta N}_{\beta_{1}^{'}\beta_{2}^{'}N^{'}L^{'}M_{L}^{'}}({\vec{r}})=
 \sum_{l sJM_{J}TM_T } \Psi^{\Delta N{lsJM_{J}TM_T}
 } _{\beta_{1}^{'}\beta_{2}^{'}N^{\prime}L^{\prime}M_{L^{\prime}}}(r)
\langle \stackrel{\symbol{94}}{\vec{r}}\mid (ls)JM_{J}TM_T\rangle.
 $$
 Here,
 $$
 \Psi^{\Delta
N{lsJM_{J}TM_T} }
_{\beta_{1}^{\prime}\beta_{2}^{\prime}N^{\prime}L^{\prime}M_{L^{\prime}}}(r)=
 \sum_{\gamma^{\prime}}
  \Psi^{^{\Delta
N}{ lsJM_{J}TM_{T}}} _{\beta_{1}^{'}\beta_{2}^{'}N^{\prime}L^{\prime}M_{L^{\prime}}
\gamma^{\prime}}(r),
 $$
  $\beta _i^{^{\prime }}=n_i^{^{\prime }}l_i^{^{\prime
}}m_{l_i}^{^{\prime }}s_i^{^{\prime }}m_{s_i}^{^{\prime }}t_i^{^{\prime
}}m_{t_i}^{^{\prime }}$,
 $\gamma^{\prime}=s^{'}\Lambda^{'}n^{'}l^{'}$ and
$$ \Psi^{^{\Delta N}{ lsJM_{J}TM_{T}}}
_{\beta_{1}^{'}\beta_{2}^{'}N^{\prime}L^{\prime}M_{L^{\prime}}
\gamma^{\prime}}({r}) = F_{\beta_{1}^{''}\beta_{2}^{''}s^{^{\prime
}}l^{^{\prime }}\Lambda ^{^{\prime }}}^{JM_JL^{^{\prime
}}M_{L^{^{\prime }}}}C_{ t_1^{^{\prime }}M_{t_1^{^{\prime
}}}t_2^{^{\prime }}m_{t_2^{^{\prime }}}}^{ Tm_{T }}a_{n^{^{\prime
}}l^{^{\prime }}N^{^{\prime }}L^{^{\prime }}}^{n_1^{^{\prime
}}l_1^{^{\prime }}n_2^{^{\prime }}l_2^{^{\prime }}\Lambda
^{^{\prime }}}(1-(-)^{l^{^{\prime }}+s^{^{\prime
}}+T})W_{ls,n^{^{\prime }}l^{^{\prime }}s^{^{\prime }}}^{\Delta
NJTM_{J}M_{T}}(r).
 $$
 The value $ W_{ls,n^{^{\prime }}l^{^{\prime }}s^{^{\prime }}}^{\Delta
NJTM_{J}M_{T}}(r)$ has the form
 $$
  W_{ls,n^{^{\prime }}l^{^{\prime }}s^{^{\prime }}}^{\Delta
NJTM_{J}M_{T}}(r)=-\int_0^\infty dr^{^{\prime }}r^{^{\prime
}2}G_l(r,r^{^{\prime }};\mu _{12},b_{12})V_{ls,l^{^{\prime
}}s^{^{\prime }}}^{JTM_{J}M_{T}}(r^{^{\prime }})R_{n^{^{\prime
}}l^{^{\prime }}}(r^{^{\prime }}), $$ $$V_{ls,l^{\prime }s^{\prime
}}^{JTM_{J}M_{T}}(r^{\prime })=\int d\hat{\vec{r}^{'}} \langle
TM_TJM_J(ls)\mid \hat{\vec{r}^{'}} \rangle V_{\Delta
N,NN}({\vec{r} }^{\prime })\langle \hat{{\vec{r}}^{\prime }}\mid
TM_TJM_J(l^{\prime }s^{\prime })\rangle. $$
 By definition
 $$
 F_{\beta_{1}^{''}\beta_{2}^{''}s^{^{\prime }}l^{^{\prime
}}\Lambda'}^{JM_JL^{^{\prime }}M_{L^{^{\prime }}}}=\sum _{m_{s'}m_{l' }M_{\Lambda'}
}
 C_{s_1^{^{\prime }}m_{s_1^{^{\prime
}}}s_2^{^{\prime }}m_{s_2^{^{\prime }}}}^{s^{^{\prime }}m_{s^{^{\prime
}}}}C_{l_1^{^{\prime }}m_{l_1^{^{\prime }}}l_2^{^{\prime }}m_{l_2^{^{\prime
}}}}^{\Lambda ^{^{\prime }}M_{\Lambda ^{^{\prime }}}}
 C_{l^{^{\prime }}m_{l^{^{\prime
}}}L^{^{\prime }}M_{L^{^{\prime }}}}^{\Lambda ^{^{\prime
}}M_{\Lambda ^{^{\prime }}}}C_{l^{^{\prime }}m_{l^{^{\prime
}}}s^{^{\prime }}m_{s^{^{\prime }}}}^{JM_J} $$ and
$\beta^{\prime\prime}_{i}=l^{\prime}_{i}m_{l^{\prime
}_{i}}s^{\prime }_{i}m_{s^{\prime }_{i}}$. The projections of the
angular momenta $ M_{J},M_{L'},m_{l_{i}'}, m_{s_{i}'}$ are fixed.
 The two-particle propagator is equal
  $$
  G_{l}(r,r^{\prime};\mu_{\Delta N},B_{\Delta N})=
  $$
  $$
   =\frac{4}{\pi}\mu_{\Delta N}B_{\Delta N}k_{l}(B_{\Delta
  N}r)i_{l}(B_{\Delta N}r^{\prime})\hspace{3cm}    r>r^{\prime}
  $$
  $$
   =\frac{4}{\pi}\mu_{\Delta N}B_{\Delta N}i_{l}(B_{\Delta
  N}r)k_{l}(B_{\Delta N}r^{\prime}) \hspace{3cm}  r<r^{\prime}
  $$
  $$
  \frac{1}{\mu_{\Delta N}}=\frac{1}{M_{\Delta}}+\frac{1}{M_{N}}
  $$
  $$
  B_{\Delta N}=\sqrt{2\mu_{\Delta N}(\frac{K^{2}}{2M}+\Delta M_{1}
  +\Delta M_{2}-E_{(\alpha\beta})}.
  $$
 Here, $ \Delta M_{1}=M_{\Delta}-M$,  $ \Delta M_{2}=M_{N}-M_{M}=0$.

 The transition potential was obtained  in the OBE approximation (one-$\pi$ and
 one-$\rho$ exchanges) and  have the form [11]
  $$
  V_{\Delta N,NN}(\vec r)= \frac{m_{\pi}}{3}V_{\Delta
  N}^{\pi}(\vec{\tau_{1}}\cdot\vec{\tau_{2}})[(\vec{\sigma_{1}}\cdot\vec{\sigma_{2}})
  (v_{0}(m_{\pi}r) +2\frac{m_{\rho}}{m_{\pi}}\frac{V_{\Delta
  N}^{\rho}}{V_{\Delta
  N}^{\pi}}v_{0}(m_{\rho}r)) +
  $$
  $$
  S_{12}(v_{2}(m_{\pi}r) -\frac{m_{\rho}}{m_{\pi}}\frac{V_{\Delta
  N}^{\rho}}{V_{\Delta
  N}^{\pi}}v_{2}(m_{\rho}r))].
  $$
   Here, the operator $S_{12}$ is
   $$
   S_{12}=\sqrt{24\pi}[\sigma_{1}^{[1]}\times \sigma_{2}^{[1]}]^{[2]}\times
   Y^{[2]}(\hat{r}]^{[0]}.
   $$
 The matrix element of the transition potential
    $$
    V_{ls,l^{'}s^{'}}^{JTM_{J}M_{T};J'T'M_{J'}M_{T'}}(r)=\int
d\hat{\vec{r}^{'}}<(l(s_{1}s_{2})s)J(t_{1}t_{2})
    T\mid \hat{\vec{r}^{'}}>V_{\Delta N,NN}(\vec{r'})
    $$
    $$
    <\hat{\vec{r}^{'}}\mid
    (l^{'}(s_{1}^{'}s_{2}^{'})s^{'})J^{'}M_{J'}(t_{1}^{'}t_{2}^{'})T^{'}M_{T'}>.
    $$
     is written as follows

     $$
      V_{ls,l^{'}s^{'}}^{JTM_{J}M_{T};J'T'M_{J'}M_{T'}}(r)=\delta_{J'J}\delta_{T'T}
      \delta_{M_{J'}M_{J}} \delta_{M_{T'}M_{T}}V_{ls,l^{'}s^{'}}^{JT}(r)
     $$
     and
    $$
     V_{ls,l^{'}s^{'}}^{JT}(r)=
      \frac{m_{\pi}}{3}V_{\Delta
     N}^{\pi}[M_{ls,l^{'}s^{'}}^{JT}(1)(
     v_{0}(m_{\pi}r) +2\frac{m_{\rho}}{m_{\pi}}\frac{V_{\Delta
     N}^{\rho}}{V_{\Delta
     N}^{\pi}}v_{0}(m_{\rho}r)) +
     $$
     $$
     M_{ls,l^{'}s^{'}}^{JT}(2)(v_{2}(m_{\pi}r) -\frac{m_{\rho}}{m_{\pi}}\frac{V_{\Delta
     N}^{\rho}}{V_{\Delta
     N}^{\pi}}v_{2}(m_{\rho}r))].
     $$
     Here,
     $$
     M_{ls,l^{'}s^{'}}^{JT}(1)=(-1)^{t_{1}+t_{2}^{'}+T}
     \left\{
      \begin{array}{c}
      Tt_{2}t_{1} \\
     1t_{1}^{'}t_{2}^{'}
     \end{array}
     \right\} \times
     $$
     $$
     <t_{1}\parallel \tau^{[1]}(1)\parallel t_{1}^{'}><t_{2}\parallel
     \tau^{[1]}(2)\parallel t_{2}^{'}>
     <s_{1}\parallel \sigma^{[1]}(1)\parallel s_{1}^{'}><s_{2}\parallel
     \sigma^{[1]}(2)\parallel s_{2}^{'}>\times
     $$
     $$
     (-1)^{s_{1}+s_{2}^{'}+s^{'}}\delta_{ll^{'}}
    \left\{
      \begin{array}{c}
      s^{'}s_{2}s_{1} \\
     1s_{1}^{'}s_{2}^{'}
      \end{array}
      \right \}
      $$
      $$
     M_{ls,l^{'}s^{'}}^{JT}(2)=
     (-1)^{t_{1}+t_{2}^{'}+T}
     \left\{
      \begin{array}{c}
      Tt_{2}t_{1} \\
     1t_{1}^{'}t_{2}^{'}
     \end{array}
     \right\} \times
      $$
      $$
     <t_{1}\parallel \tau^{[1]}(1)\parallel t_{1}^{'}><t_{2}\parallel
     \tau^{[1]}(2)\parallel t_{2}^{'}>
     <s_{1}\parallel \sigma^{[1]}(1)\parallel s_{1}^{'}><s_{2}\parallel
     \sigma^{[1]}(2)\parallel s_{2}^{'}>\times
     $$
     $$
     \sqrt{30}(-1)^{J+s}\hat{s^{'}}\hat{s}\hat{l}\hat{l^{'}}
     $$
     $$
     \left (
      \begin{array}{c}
      l^{'}2l \\
      000
     \end{array}
     \right )
     \left\{
      \begin{array}{c}
      Jls \\
      2s^{'}l^{'}
     \end{array}
     \right\}
     \left\{
      \begin{array}{c}
      s_{1}s_{1}^{'}1 \\
      s_{2}s_{2}^{'}1\\
      ss^{'}2
      \end{array}
      \right\}
      $$
      and
      $$
      v_{l}(r)=\frac{2}{\pi}(-1)^{l_{1}+l_{2}}\frac{\Lambda^{4}}{(\Lambda^{2}-m^{2})
      ^{2}}
      [k_{l}(mr)-(\frac{\Lambda}{m})^{l_{1}+l_{2}+1}[k_{1}(\Lambda
      r)+
      $$
      $$
      \frac{\Lambda^{2}-m^{2}}{2\Lambda^{2}}(\Lambda
      rk_{l-1}(\Lambda r)-(l_{1}+l_{2}-l)k_{l}(\Lambda r)]].
      $$

 Here,
$$
V _ {\Delta N} ^ {\pi} = \frac {f _ {\Delta N\pi} f _ {NN\pi}} {4\pi},
$$
$$
V _ {\Delta N} ^ {\rho} = \frac {f _ {\Delta N\rho} f _ {NN\rho}} {4\pi},
$$
$\Lambda$ is the regularization parameter.
 For the constants we take the following values
$$
\frac {V _ {\Delta N} ^ {\rho}} {V _ {\Delta N} ^ {\pi}} = 3.9953[14].
$$
and
$$
V _ {\Delta N} ^ {\pi} = 0.17 [15].
$$
 Therefore, the values written above are
 $$
  V_{ls,l^{'}s^{'}}^{JTM_{J}M_{T}}(r)= \delta_{JJ}\delta_{TT}
  \delta_{M_{J}M_{J}} \delta_{M_{T}M_{T}}V_{ls,l^{'}s^{'}}^{JT}(r),
 $$
 $$
 W_{ls,l^{'}s^{'}}^{JTM_{J}M_{T}}(r)= \delta_{JJ}\delta_{TT}
 \delta_{M_{J}M_{J}} \delta_{M_{T}M_{T}}W_{ls,l^{'}s^{'}}^{JT}(r).
$$ Here,
 $$
 W_{ls,n^{^{\prime }}l^{^{\prime }}s^{^{\prime
}}}^{\Delta NJT}(r)=-\int_0^\infty dr^{^{\prime }}r^{^{\prime
}2}G_l(r,r^{^{\prime }};\mu _{12},b_{12})V_{ls,l^{^{\prime
}}s^{^{\prime }}}^{JT}(r^{^{\prime }})R_{n^{^{\prime }}l^{^{\prime
}}}(r^{^{\prime }}). $$
 b) After transition from the coordinates $ \vec{r_{1}} $, $ \vec{r_{2}} $ to relative $
\vec{r}$ and  c.m. $ \vec{R} $ coordinates the wave function of
the  $ \Delta N $ system in the case the jj- coupling can be
written as follows
 $$\Psi _{\alpha _1^{^{\prime }}\alpha
_2^{^{\prime }}}^{\Delta N}( \vec{r_{1}},
\vec{r_{2}})=\sum_{\alpha}\Psi^{\Delta N} _{\alpha _1^{\prime
}\alpha _2^{\prime }\alpha } (\vec{r}) \Phi _{\alpha} (\vec{R}),
$$
 where $\alpha
=N^{^{\prime }}L^{^{\prime }}M_{L^{^{\prime }}}$ and
$$\Psi^{\Delta N}
_{\alpha\alpha _1^{\prime }\alpha _2^{\prime }} (\vec{r})= \sum _{m_{l_1^{^{\prime
}}}m_{l_2^{^{\prime}}}}\sum _{m_{s_1^{^{\prime }}}m_{s_2^{^{\prime }}}} C_{
l_1^{^{\prime }}m_{l_1^{^{\prime }}}s_1^{^{\prime }}m_{S_1^{^{\prime }}}}^{
j_1^{^{\prime }}m_1^{^{\prime }}} C_{ l_2^{^{\prime }}m_{l_2^{^{\prime
}}}s_2^{^{\prime }}m_{S_2^{^{\prime }}}}^{ j_2^{^{\prime }}m_2^{^{\prime }}} \times
 \Psi _{\beta _1^{^{\prime }}\beta
_2^{^{\prime }}\alpha }^{\Delta N}(\vec{r}).
$$
 Let us define
 $$
 F_{\alpha_1^{^{\prime
\prime }}\alpha_2^{^{\prime \prime}}s^{^{\prime }}l^{^{\prime
}}\Lambda ^{^{\prime }}}^{JM_JL^{^{\prime }}M_{L^{^{\prime }}}}
=\sum _{m_{l_1^{^{\prime }}}m_{l_2^{^{\prime }}}m_{s_1^{^{\prime
}}}m_{s_2^{^{\prime }}}}C_{l_1^{^{\prime }}m_{l_1^{^{\prime
}}}s_1^{^{\prime }}m_{s_1^{^{\prime }}}}^{j_1^{^{\prime
}}m_{j_1^{^{\prime }}}}C_{l_2^{^{\prime }}m_{l_2^{^{\prime
}}}s_2^{^{\prime }}m_{s_2^{^{\prime }}}}^{j_2^{^{\prime
}}m_{j_2^{^{\prime }}}} $$
 $$
 F_{\beta_1^{^{\prime
\prime }}\beta_2^{^{\prime \prime}}s^{^{\prime }}l^{^{\prime }}\Lambda ^{^{\prime
}}}^{JM_JL^{^{\prime }}M_{L^{^{\prime }}}}, $$
 where $\alpha_i^{^{\prime
\prime }}=l_i^{^{\prime }}j_i^{\prime}m_{j_i^{^{\prime }}}s_i^{^{\prime }}$.
 Then, the wave function of relative motion of the $ \Delta N $ system
 can be written as follows
$$ \Psi^{\Delta
N}_{{\alpha_{1}^{\prime}\alpha_{2}^{\prime}}N^{\prime}L^{\prime}M_{L^{\prime}}}(
\vec{r} )=\sum_{\gamma^{\prime},JM_{J}TM_{T}ls} \Psi^{\Delta N
lsJM_{J}TM_{T}}
_{{\alpha_{1}^{\prime}\alpha_{2}^{\prime}}N^{\prime}L^{\prime}M_L^{\prime}
;\gamma^{\prime}}(r) <\hat{\vec{r}}\mid (ls)JM_{J} TM_{T}> , $$
where $\gamma^{\prime}\equiv s^{'}\Lambda^{'}n^{'}l^{'}.$ It its
turn, the radial wave function is given by
 $$
  \Psi^{\Delta
N lsJM_{J}TM_{T}}
_{{\alpha_{1}^{\prime}\alpha_{2}^{\prime}}N^{\prime}L^{\prime}M_L^{\prime}
;\gamma^{\prime}}(r)= F_{\alpha_1^{^{\prime \prime
}}\alpha_2^{^{\prime \prime}}s^{^{\prime }}l^{^{\prime }}\Lambda
^{^{\prime }}}^{JM_JL^{^{\prime }}M_{L^{^{\prime }}}} C_{
t_1^{^{\prime }}m_{t_1^{^{\prime }}}t_2^{^{\prime
}}m_{t_2^{^{\prime }}}}^{ Tm_{T }}a_{n^{^{\prime }}l^{^{\prime
}}N^{^{\prime }}L^{^{\prime }}}^{n_1^{^{\prime }}l_1^{^{\prime
}}n_2^{^{\prime }}l_2^{^{\prime }}\Lambda ^{^{\prime
}}}(1-(-)^{l^{^{\prime }}+S^{^{\prime }}+T})W_{ls,n^{^{\prime
}}l^{^{\prime }}s^{^{\prime }}}^{\Delta NJTM_{J}M_{T}}(r). $$ It
can be show that
 $$
 F_{\alpha_1^{^{\prime \prime
}}\alpha_2^{^{\prime \prime}}s^{^{\prime }}l^{^{\prime }}\Lambda ^{^{\prime
}}}^{JM_JL^{^{\prime }}M_{L^{^{\prime }}}}
=(-1)^{l_{1}^{'}+l_{2}^{'}+s_{1}^{'}-s_{2}^{'}+j_{1}^{'}+j_{2}^{'}+\Lambda^{'}+2l'-
2s'-2J-2L'}\hat{j_{1}^{'}}\hat{j_{2}^{'}}\hat{s'}\hat{\Lambda'}^{2}\hat{J} {\sum
}_{x\xi }(-1)^{x-\xi}(2x+1)$$

$$\left(
\begin{array}{c}
j_{1}^{^{\prime }}xj_{2}^{^{\prime }} \\
m_{j_{1}^{^{\prime }}}\xi m_{j_{2}^{^{\prime }}}
\end{array}
\right) \left(
\begin{array}{c}
L^{^{\prime }}xJ \\
M_{L^{^{\prime }}}\xi M_{J}
\end{array}
\right) \left\{
\begin{array}{c}
L^{^{\prime }}xJ \\
s^{^{\prime }}l^{^{\prime }}\Lambda ^{^{\prime }}
\end{array}
\right\} \left\{
\begin{array}{c}
\begin{array}{c}
j_{1}^{^{\prime }}xj_{2}^{^{\prime }} \\
l_{1}^{^{\prime }}\Lambda ^{^{\prime }}l_{2}^{^{\prime }}
\end{array}
\\
s_{1}^{^{\prime }}s^{^{\prime }}s_{2}^{^{\prime }}
\end{array}
\right\}.
 $$
Here, $\left\{\right\}$ are the 6j- 9j- symbols.

c)  By definition, the norm of the wave function of the $ \Delta $ N system  is
$$
N=\sum_{\delta_{1}^{\prime}\delta_{2}^{\prime}}\int\Psi^{+\Delta
N}_{\delta_{1}^{\prime}\delta_{2}^{\prime}}(\vec{r}_{1}^{\prime}\vec{r}_{2}^{\prime})
\Psi^{\Delta
N}_{\delta_{1}^{\prime}\delta_{2}^{\prime}}(\vec{r}_{1}^{\prime}\vec{r}_{2}^{\prime})
d\vec{r}_{1}^{\prime}d\vec{r}_{2}^{\prime}.
$$
The indices $ \delta _ {i} ^ {\prime} $ (i = 1,2) are identified with the indices $
\beta _ {i} ^ {\prime} $ in the case of the ls-coupling and the indices $ \alpha _
{i} ^ {\prime} $ in the case the jj- coupling. Here, the sum over the spin and
isospin variables is meant. After transition to relative coordinate and
 c. m. coordinate we have
$$
N= \sum_{\delta_{1}^{\prime}\delta_{2}^{\prime}}\int d\vec{r}
\sum_{N^{\prime}L^{\prime}M_{L^{\prime}}}\Psi^{\star\Delta
N}_{\delta_{1}^{\prime}\delta_{2}^{\prime}N^{\prime}L^{\prime}M_{L^{\prime}}}(\vec{r})
\Psi^{\Delta
N}_{\delta_{1}^{\prime}\delta_{2}^{\prime}N^{\prime}L^{\prime}M_{L^{\prime}}}(\vec{r}).
$$
For transformation of the last ratio we shall take  the formulae mentioned above.
Performing the angular variables integration, we have
$$
N= \sum_{\delta_{1}^{\prime}\delta_{2}^{\prime}}
\sum_{N^{\prime}L^{\prime}M_{L^{\prime}}}
\sum_{lsJM_{J}TM_{T}}\sum_{n^{\prime}l^{\prime}s^{\prime}\Lambda^{\prime}}
\sum_{\tilde{n^{\prime}}\tilde{l^{\prime}}\tilde{s^{\prime}}\tilde{\Lambda^{\prime}}}
$$
$$
F_{\delta_{1}^{''}\delta_{2}^{''}s^{^{\prime }}l
 ^{^{\prime
 }}\Lambda ^{^{\prime }}}^{JM_JL^{^{\prime }}M_{L^{^{\prime }}}}
 F_{\delta_{1}^{''}\delta_{2}^{''}{\tilde s}^{\prime }{\tilde l}
 ^{\prime
 }{\tilde \Lambda} ^{\prime }}^{{ J}{ M}_J{ L}^{\prime}{M}
 _{L^{\prime }}}
 C_{
t_1^{^{\prime }}m_{t_1^{^{\prime }}}t_2^{^{\prime
}}m_{t_2^{^{\prime }}}}^{ Tm_{T }}C_{ t_1^{^{\prime
}}m_{t_1^{^{\prime }}}t_2^{^{\prime }}m_{t_2^{^{\prime }}}}^{
Tm_{T }}a_{n^{^{\prime }}l^{^{\prime }}N^{^{\prime }}L^{^{\prime
}}}^{n_1^{^{\prime }}l_1^{^{\prime }}n_2^{^{\prime }}l_2^{^{\prime
}}\Lambda ^{^{\prime }}} a_{{\tilde n}^{^{\prime }}{\tilde
l}^{^{\prime }}N^{^{\prime }}L^{^{\prime }}}^{n_1^{^{\prime
}}l_1^{^{\prime }}n_2^{^{\prime }}l_2^{^{\prime }}{\tilde\Lambda}
^{^{\prime }}} $$ $$ (1-(-)^{l^{^{\prime }}+s^{^{\prime
}}+T})(1-(-)^{{\tilde l}^{^{\prime }}+{\tilde s}^{^{\prime }}+T})
$$
 $$ \int W_{ls,n^{^{\prime }}l^{^{\prime }}s^{^{\prime
}}}^{\star\Delta NJTM_{J}M_{T}}(r) W_{ls,{\tilde n}^{^{\prime
}}{\tilde l}^{^{\prime }}{\tilde s}^{^{\prime }}}^{\Delta
NJTM_{J}M_{T}}(r)r^2 dr.
 $$
  Using  the formulae for F  mentioned
above, we have the following expressions for
 the norm in case of the ls- coupling
$$
N=\sum_{n^{\prime}_{1}n^{\prime}_{2}l^{\prime}_{1}l^{\prime}_{2}}
\sum_{N^{\prime}L^{\prime}M_{L^{\prime}}} \sum_{lsJT} \sum_{s^{\prime}l^{\prime}}
\sum_{n^{\prime}\tilde{n^{\prime}}}
\frac{\hat{\Lambda^{\prime}}^{2}\hat{J}^{2}\hat{T}^{2}}
 {\hat{l^{\prime}}^{2}}2(1-(-1)^{l^{\prime}+s^{\prime}+T})
 $$
 $$
 a_{n^{^{\prime }}l^{^{\prime }}N^{^{\prime }}L^{^{\prime }}}^{n_1^{^{\prime
}}l_1^{^{\prime }}n_2^{^{\prime }}l_2^{^{\prime }}\Lambda
^{^{\prime }}} a_{{\tilde n}^{^{\prime }}l^{^{\prime }}N^{^{\prime
}}L^{^{\prime }}}^{n_1^{^{\prime }}l_1^{^{\prime }}n_2^{^{\prime
}}l_2^{^{\prime }}\Lambda ^{^{\prime }}} \int W_{ ls,n^{^{\prime
}}l^{^{\prime }}s^{^{\prime }}}^{\star \Delta NJT}(r)
W_{ls,{\tilde n}^{^{\prime }}{l}^{^{\prime }}{ s}^{^{\prime
}}}^{\Delta NJT}(r)r^{2}dr. $$
 In the case of the jj- coupling the
norm of the wave function of the isobar configuration can be
written as follows
 $$
  N=
\sum_{n_{1}'l_{1}'j_{1}'n_{2}'l_{2}'j_{2}'}\int r^{2}dr
\sum_{N^{\prime}L^{\prime}}\sum_{lsJT}
\sum_{n^{'}l^{'}s^{'}\Lambda^{'}}\sum_{\tilde{n^{'}}\tilde{l^{'}}\tilde{s^{'}}
\tilde{\Lambda^{'}}} (-1)^{a+ {\tilde
a}}\hat{j_{1}'^{2}}\hat{j_{2}'^{2}}{\hat{\tilde{\Lambda'^2}}}{\hat{\Lambda'^{2}}}
\hat{{J}}^{2}\sum _{x}\hat{x}^{2}\{j_{1}^{\prime }j_{2}^{\prime
}x\} $$
\bigskip
$$\left\{
\begin{array}{c}
L^{\prime}xJ \\
s^{\prime }l^{\prime }\Lambda ^{\prime }
\end{array}
\right\} \left\{
\begin{array}{c}
{ L}^{\prime }x{ J} \\
{\tilde s}^{\prime }{\tilde l}^{\prime } {\tilde \Lambda }^{\prime }
\end{array}
\right\} \left\{
\begin{array}{c}
\begin{array}{c}
j_{1}^{\prime }xj_{2}^{\prime } \\
l_{1}^{\prime }\Lambda ^{\prime }l_{2}^{\prime }
\end{array}
\\
s_{1}^{\prime }s^{\prime }s_{2}^{\prime }
\end{array}
\right\} \left\{
\begin{array}{c}
\begin{array}{c}
j_{1}^{\prime }xj_{2}^{\prime } \\
l_{1}^{\prime }{\tilde \Lambda }^{\prime }l_{2}^{\prime }
\end{array}
\\
s_{1}^{\prime }\tilde{{s}^{\prime }}s_{2}^{\prime }
\end{array}%
\right\} .
$$
$$
 a_{n^{^{\prime }}l^{^{\prime }}N^{^{\prime }}L^{^{\prime }}}^{n_1^{^{\prime
}}l_1^{^{\prime }}n_2^{^{\prime }}l_2^{^{\prime }}\Lambda
^{^{\prime }}}a_{\tilde{n^{^{\prime }}}\tilde{l^{^{\prime
}}}N^{^{\prime }}L^{^{\prime }}}^{n_1^{^{\prime }}l_1^{^{\prime
}}n_2^{^{\prime }}l_2^{^{\prime }}\Lambda ^{^{\prime }}} $$ $$
(1-(-)^{l^{^{\prime }}+s^{^{\prime
}}+T})(1-(-)^{\tilde{l^{^{\prime }}}+\tilde{s^{^{\prime }}}+T})
 $$
 $$
W_{ls,n^{^{\prime }}l^{^{\prime }}s^{^{\prime }}}^{\Delta NJT}(r)
W_{ls,\tilde{n^{^{\prime }}}\tilde{l^{^{\prime
}}}\tilde{s^{^{\prime }}}}^{\Delta NJT}(r). $$

\textbf{Appendix 2.}\\
With the help of the graphic method of summation over the projections of the angular
momenta [13] it is possible to receive  the following expression in the case of the
ls- coupling
 $$
\sum_{m^{\prime}_{l_{1}}m^{\prime}_{l_{2}}m^{\prime}_{s_{1}}
 m^{\prime}_{s_{2}}}F_{\beta_{1}^{''}\beta_{2}^{''}s^{^{\prime }}l
 ^{^{\prime
 }}\Lambda ^{^{\prime }}}^{JM_JL^{^{\prime }}M_{L^{^{\prime }}}}
 F_{\beta_{1}^{''}\beta_{2}^{''}{\tilde s}^{\prime }{\tilde l}
 ^{\prime
 }{\tilde \Lambda} ^{\prime }}^{{\tilde J}{\tilde M}_J{\tilde L}^{\prime}{\tilde M}
 _{L^{\prime }}}=
 $$
 $$
 \sum_{xm_{x}}\beta(x)(-1)^{x-m_{x}}\hat{x}^{2}(-1)^{{\tilde L}^{\prime }-M_{{\tilde
L}^{\prime }}}(-1)^{\tilde J-{\tilde M_{J}}}
 $$
 $$
\left(
\begin{array}{c}
{\tilde L}^{\prime }xL^{\prime}  \\
-M_{\tilde {L^{\prime}}}m_{x}M_{L^{\prime}}
\end{array}
\right)
 \left(
\begin{array}{c}
 \tilde JxJ\\
-M_{\tilde J}-m_{x}M_{J}
\end{array}
\right),
 $$
 where
 $$
 \beta(x)=\delta_{s^{\prime}{\tilde s}^{\prime}}
 \delta_{\Lambda^{\prime}{\tilde \Lambda}^{\prime}}
 (-1)^{-l^{\prime}-{\tilde l}^{\prime}-s^{\prime}-\Lambda^{\prime}-2L^{\prime}
 +{\tilde
 L}^{\prime}-J}{\hat {\Lambda^\prime}}^{2}\hat{J}\hat{{\tilde
 J}}
 \left\{
\begin{array}{c}
{\tilde J}x J \\
 l^{\prime}s^{\prime} {\tilde l}^{\prime}
\end{array}
\right\} \left\{
\begin{array}{c}
L^{\prime} x{\tilde L}^{\prime}  \\
{\tilde l}^{\prime}\Lambda^{\prime}l^{\prime}
\end{array}
\right\}.
 $$
 The analogous summation  gives in the case of the  jj- coupling
$$
  \sum_{m_{j_{1}^{\prime
}}m_{j_{2}^{\prime }}}  F_{\alpha_{1}^{''}\alpha_{2}^{''}s^{'}l
 ^{'}\Lambda ^{'}}^{JM_JL^{'}M_{L^{'}}}
 F_{\alpha_{1}^{''}\alpha_{2}^{''}{\tilde s}^{\prime }{\tilde l}
 ^{\prime
 }{\tilde \Lambda} ^{\prime }}^{{\tilde J}{\tilde M}_J{\tilde L}^{\prime}{\tilde M}
 _{L^{\prime }}}=
 $$
 $$
(-1)^{a+ {\tilde
a}}\hat{j_{1}'^{2}}\hat{j_{2}'^{2}}{\hat{\tilde{\Lambda'^2}}}{\hat{\Lambda'^{2}}}
\hat{\tilde{J}}\hat{J}\sum _{xm_{x}}(-1)^{2x-2m_{x}}(2x+1)\{j_{1}^{\prime
}j_{2}^{\prime }x\}
\left(
\begin{array}{c}
L^{\prime }xJ \\
M_{L^{\prime }}\xi M_{J}
\end{array}
\right)
 \left(
 \begin{array}{c}

 {\tilde L}^{\prime }x{\tilde J}
\\
{\tilde M}_{L^{\prime }}\xi{\tilde M}_{J}
\end{array}
\right) $$
\bigskip
$$\left\{
\begin{array}{c}
L^{\prime}xJ \\
s^{\prime }l^{\prime }\Lambda ^{\prime }
\end{array}
\right\} \left\{
\begin{array}{c}
{\tilde L}^{\prime }x{\tilde J} \\
{\tilde s}^{\prime }{\tilde l}^{\prime } {\tilde \Lambda }^{\prime }
\end{array}
\right\} \left\{
\begin{array}{c}
\begin{array}{c}
j_{1}^{\prime }xj_{2}^{\prime } \\
l_{1}^{\prime }\Lambda ^{\prime }l_{2}^{\prime }
\end{array}
\\
s_{1}^{\prime }s^{\prime }s_{2}^{\prime }
\end{array}
\right\} \left\{
\begin{array}{c}
\begin{array}{c}
j_{1}^{\prime }xj_{2}^{\prime } \\
l_{1}^{\prime }{\tilde \Lambda }^{\prime }l_{2}^{\prime }
\end{array}
\\
s_{1}^{\prime }{s}^{\prime }s_{2}^{\prime }
\end{array}%
\right\} .
$$
Here,
$a={l_{1}^{'}+l_{2}^{'}+s_{1}^{'}-s_{2}^{'}+j_{1}^{'}+j_{2}^{'}+\Lambda^{'}+2l'-
2s'-2J-2L'}$,
$\tilde{a}={l_{1}^{'}+l_{2}^{'}+s_{1}^{'}-s_{2}^{'}+j_{1}^{'}+j_{2}^{'}+
\tilde{\Lambda^{'}}+2l'- 2s'-2\tilde{J}-2\tilde{L'}}.$

 In the case of the ls- coupling it is possible to show that
$$
\sum_{m^{\prime}_{l_{1}}m^{\prime}_{l_{2}}m^{\prime}_{s_{1}}
 m^{\prime}_{s_{2}}}\sum_{M_{L^{\prime}}{\tilde M}
 _{L^{\prime }}M_{J}{\tilde M}_J}
 F_{\beta_{1}^{''}\beta_{2}^{''}s^{^{\prime }}l
 ^{^{\prime
 }}\Lambda ^{^{\prime }}}^{JM_JL^{^{\prime }}M_{L^{\prime }}}
 F_{\beta_{1}^{''}\beta_{2}^{''}{\tilde s}^{\prime }{\tilde l}
 ^{\prime\
 }{\tilde \Lambda} ^{\prime }}^{{\tilde J}{\tilde M}_J{\tilde L}^{\prime}{\tilde M}
 _{L^{\prime }}}
 A_{{\tilde L}^{\prime }{\tilde M}^{\prime };( {\tilde l}{\tilde
s}){\tilde J}{\tilde M}_{J} {\tilde T}{\tilde M}_{T}}^{L^{\prime }M_{L^{\prime
}};(ls)JM_{J}TM_{T}}(k,p,P)=
$$
$$
\sum_{\lambda}\frac{\alpha(\lambda)\beta(\lambda)}{2\lambda+1}(-1)^{\lambda+J-{\tilde
J}+2L^{\prime}}\{{\tilde J}J \lambda\}\{{\tilde L}^{\prime }L^{\prime}\lambda\}
$$
$$
\int q^{2}dqj_{0}(kq)j_{\lambda }(pq)j_{\lambda }(\frac{M_{1}}{M}Pq),
$$
were $ M=M_{1}+ M_{2}$.
 In the case of the jj- coupling we have
$$
 \sum_{m_{j_{1}}'m_{j_{2}}'}\sum_{ M_{J}{\tilde M_{J}}M_{L^{\prime }}
 {\tilde M_{L^{\prime }}}}
  F_{\alpha_{1}^{''}\alpha_{2}^{''}s^{'}l
 ^{'}\Lambda ^{'}}^{JM_JL^{'}M_{L^{'}}}
 F_{\alpha_{1}^{''}\alpha_{2}^{''}{\tilde s}^{\prime }{\tilde l}
 ^{\prime
 }{\tilde \Lambda} ^{\prime }}^{{\tilde J}{\tilde M}_J{\tilde L}^{\prime}{\tilde M}
 _{L^{\prime }}}
 A_{{\tilde L}^{\prime
}{\tilde M}^{\prime };( {\tilde l}{\tilde s}){\tilde J}{\tilde M}_{J} {\tilde
T}{\tilde M}_{T}}^{L^{\prime }M_{L^{\prime }};(ls)JM_{J}TM_{T}}(k,p,P)=
 $$
 $$
 (-1)^{a+
{\tilde a}}(-1)^{\tilde{l}+s+2J+\tilde{J}}\hat{j_{1}'^{2}}\hat{j_{2}'^{2}}
{\hat{\tilde{\Lambda'^2}}} {\hat{\Lambda'^{2}}}
\hat{\tilde{J}^{2}}\hat{J}^{2}\delta_{T{\tilde T}}\delta_{M_{T}{\tilde
M}_{T}}\delta_{s\tilde s}\frac{2}{\pi} {\hat l}{\hat L}^\prime
$$
$$
 C^{\tilde{L'}0}_{L'0\lambda 0}
C^{\tilde{l}0}_{l0\lambda 0}
 \sum _{x\lambda}(-1)^{\lambda+x+L'}(2\lambda+1)(2x+1)\{j_{1}^{\prime
}j_{2}^{\prime }x\}
 \left \{
\begin{array}{c}
\lambda \tilde{J}J \\
sl\tilde{l}
\end{array}
\right \}
 \left \{
 \begin{array}{c}

 J{\tilde J}\lambda
\\
\tilde{L'}L'x
\end{array}
\right \} $$
\bigskip
$$\left\{
\begin{array}{c}
L^{\prime}xJ \\
s^{\prime }l^{\prime }\Lambda ^{\prime }
\end{array}
\right\} \left\{
\begin{array}{c}
{\tilde L}^{\prime }x{\tilde J} \\
{\tilde s}^{\prime }{\tilde l}^{\prime } {\tilde \Lambda }^{\prime }
\end{array}
\right\} \left\{
\begin{array}{c}
\begin{array}{c}
j_{1}^{\prime }xj_{2}^{\prime } \\
l_{1}^{\prime }\Lambda ^{\prime }l_{2}^{\prime }
\end{array}
\\
s_{1}^{\prime }s^{\prime }s_{2}^{\prime }
\end{array}
\right\} \left\{
\begin{array}{c}
\begin{array}{c}
j_{1}^{\prime }xj_{2}^{\prime } \\
l_{1}^{\prime }{\tilde \Lambda }^{\prime }l_{2}^{\prime }
\end{array}
\\
s_{1}^{\prime }{s}^{\prime }s_{2}^{\prime }
\end{array}%
\right\}
 $$
 $$
 \int q^{2}dqj_{0}(kq)j_{\lambda }(pq)j_{\lambda }(\frac{M_{1}}{M}Pq).
 $$
Taking into account the  formulae above, we can write  the  momentum disributrion in
the case of the ls- coupling as follows
$$
\Delta_{\overline {\beta} _ {1} ^ {\prime}\overline {\beta} _ {1} ^ {\prime}}=
\sum_{\lambda}\sum_{N^{'}L^{'}\gamma^{\prime}}\sum_{\tilde{N^{'}}
          \tilde{L^{'}}{\tilde\gamma}^{\prime} }\sum_{lsJT}\sum_{\tilde{l}
\tilde{s}\tilde{J}}M^{slJ \tilde s\tilde l{\tilde J}
T}_{N^{\prime}L^{\prime}\gamma^{\prime};
        {\tilde N}^{\prime} {\tilde
        L}^{\prime}{{\tilde\gamma}^{\prime}}}(\lambda)
$$
$$\int q^{2}dqj_{0}(kq)\int P^{2}dPj_{\lambda }(\frac{M_{1}}{M}Pq)
R_{N^{\prime }L^{\prime }}(P)R_{{\tilde N}^{\prime } {\tilde L}^{\prime }}(P)\int
p^{2}dp j_{\lambda }(pq)W_{ls,n^{\prime }l^{\prime }s^{\prime }}^{\Delta
NJT}(p)W_{{\tilde l}s,{\tilde n}^{\prime }{\tilde l}^{\prime }{\tilde s}^{\prime
}}^{\Delta N{\tilde J}T}(p),
$$
where
$$
M^{slJ \tilde s\tilde l{\tilde J}T}_{N^{\prime}L^{\prime}\gamma^{\prime};
        {\tilde N}^{\prime} {\tilde
        L}^{\prime}{\tilde\gamma^{\prime}}}(\lambda)=(-1)^{s+\tilde l+
        {\tilde L}^{\prime}+J
        -l^{\prime}-{\tilde l}^{\prime}-s^{\prime}-\Lambda^{\prime}}
(1-(-1)^{l^\prime +s^\prime +T})(1-(-1)^{{\tilde l}^\prime+{\tilde
s}^\prime+T})\hat{T}^{2}
$$
$$
\frac{2}{\pi}\delta_{s^{\prime}{\tilde s}^{\prime}}\delta_{\Lambda^{\prime}{\tilde
\Lambda}^{\prime}}\delta_{s {\tilde s}}\hat{l}\hat{J}^{2}\hat{\tilde
J}^{2}\hat{\Lambda^{\prime}}^{2}\hat {L^{\prime}}\hat{\lambda}^{2}C^{{\tilde
L}^{\prime }0}_{L^{\prime }0 \lambda 0}C^{{\tilde l} 0}_{l0 \lambda 0}
$$
$$
a_{n^{\prime }l^{\prime }
 N^{\prime}L^{\prime }}^{n_1^{\prime }l_1^{\prime }n_2^{\prime
}l_2^{\prime }\Lambda ^{\prime }}a_{{\tilde n}^{\prime }{\tilde l}^{\prime }
 {\tilde N}^{\prime}{\tilde L}^{\prime }}^{n_1^{\prime }l_1^{\prime }n_2^{\prime
}l_2^{\prime }{\tilde \Lambda} ^{\prime }}
 N^{slJ{\tilde l}{\tilde
J}}_{s^{\prime}l^{\prime}\Lambda^{\prime}{\tilde l}^{\prime}}(\lambda)
$$
and
$$
N^{slJ{\tilde l}{\tilde J}}_{s^{\prime}l^{\prime}\Lambda^{\prime}{\tilde
l}^{\prime}L^{\prime}{\tilde L}^{\prime}}(\lambda)=\{{\tilde J}J \lambda\}\{{\tilde
L}^{\prime }L^{\prime}\lambda\}\left\{
\begin{array}{c}
{\tilde J} J\lambda \\
 l^{\prime} {\tilde l}^{\prime}s^{\prime}
\end{array}
\right\} \left\{
\begin{array}{c}
{\tilde L}^{\prime}L^{\prime} \lambda  \\
l^{\prime}{\tilde l}^{\prime}\Lambda^{\prime}
\end{array}
\right\}\left\{
\begin{array}{c}
{\tilde J} J\lambda \\
 l {\tilde l}s
\end{array}
\right\},
$$
$$
\overline {\beta} _ {i} ^ {\prime}=n^{\prime}_{i}l^{\prime}_{i} s^{\prime}_{i}
 t^{\prime}_{i},\gamma^{\prime}=n^{\prime}l^{\prime}s^{\prime}\Lambda^{\prime}.
$$
In the final expression there is the sum over the  main quantum numbers, and the
angular momenta dependent from the initial fixed $ \beta _ {i} ^ {\prime}$
$$N^{'}L^{'}s^{\prime}l^{\prime}n^{\prime}\Lambda^{\prime} {\tilde
N}^{'}{\tilde L}^{'}{\tilde s}^{\prime}{\tilde l}^{\prime}{\tilde
n}^{\prime}{\tilde\Lambda}^{\prime}.
$$
Besides, there is the sum over the final angular momenta
$$
slJ{\tilde s}{\tilde l}{\tilde J}\lambda T.
$$
The limitations on the angular momenta follow from the 6j- symbols, which enter in
expression for the constant  $N^{slJ{\tilde l}{\tilde
J}}_{s^{\prime}l^{\prime}\Lambda^{\prime}{\tilde l}^{\prime}L^{\prime}{\tilde
L}^{\prime}}(\lambda)$. Fom them, the following rules of the vector addition for the
angular momenta should be fulfiled: $(\tilde JJ\lambda),(\lambda l^{\prime}{\tilde
l}^{\prime}),(Jl^{\prime}s^{\prime}),(s^{\prime}{\tilde J}{\tilde l}^{\prime})$ and
$({\tilde L}^{\prime}L^{\prime}\lambda),(\lambda l^{\prime}{\tilde
l}^{\prime}),(L^{\prime}l^{\prime}\Lambda^{\prime}),(\Lambda^{\prime}{\tilde
L}^{\prime}{\tilde l}^{\prime})$. At last, from the third the 6j- symbol we have
$(\tilde JJ\lambda),(\lambda l{\tilde l},(Jls),(s{\tilde J}{\tilde l})$. There are
also limitations on the final angular momenta $ s, l, J $  following from the
transition potential. The first term of the matrix element of the transition
potential is different from zero only, if the triangle relations $
(s^{\prime}s_{2}s_{1)},(s_{1}1s_{1}^{\prime}),(s_{2}1s_{2}^{\prime}),
(s_{2}^{\prime}s^{\prime}s_{1}^{\prime})$ are fulfiled and $ l = l ^ {\prime} $.
Second term in the matrix element of transition potential is different from zero
only, if the limitations on the angular momenta $ (J l s), (s2s ^ {\prime}), (l2s ^
{\prime}), (l ^ {\prime} Js ^ {\prime}), (l ^ {\prime} 2l), (s _ {1} s _ {2} s), (s
_ {1} s _ {1} ^ {\prime} 1), (s _ {1} s _ {2} ^ {\prime} s ^ {\prime}), (s _ {2} s _
{2} ^ {\prime} 1) $ following from the 3jm -, 6j -and 9j- symbols are fulfiled.

For the case of the jj- coupling  the momentum distribution of the $ \Delta $-
isobars can be written
$$
 \Delta _{\overline {\alpha} _{1}^{\prime  }\overline {\alpha} _{2}^{\prime
}}(k)= \sum_{{N^\prime L^\prime n^\prime l^\prime s^\prime \Lambda^\prime}; {{\tilde
N}^\prime {\tilde L}^\prime {\tilde n}^\prime {\tilde l}^\prime {\tilde s}^\prime
{\tilde \Lambda}^\prime}}\sum_{\lambda lsJ\tilde l\tilde JT} M_{N^\prime L^\prime
n^\prime l^\prime s^\prime \Lambda^\prime; {\tilde N}^\prime {\tilde L}^\prime
{\tilde n}^\prime {\tilde l}^\prime {\tilde s}^\prime {\tilde
\Lambda}^\prime}^{\lambda lsJ\tilde l\tilde JT} \int q^2dq j_{0}(kq)
$$
$$
\int P^2dP j_{\lambda}(\frac{M_{1}}{M}Pq) R_{N^{\prime }L^{\prime }}(P)R_{{\tilde
N}^{\prime } {\tilde L}^{\prime }}(P) \int p^2 dp j_{\lambda}(pq)W_{ls,n^{\prime
}l^{\prime }s^{\prime }}^{\Delta NJT}(p)W_{{\tilde l}s,{\tilde n}^{\prime }{\tilde
l}^{\prime }{\tilde s}^{\prime }} ^{\Delta N{\tilde J}T}(p).
$$
Here,
$$
M_{N^\prime L^\prime n^\prime l^\prime s^\prime \Lambda^\prime;
{\tilde N}^\prime {\tilde L}^\prime {\tilde n}^\prime {\tilde l}^\prime {\tilde
s}^\prime {\tilde \Lambda}^\prime}^{\lambda lsJ\tilde l\tilde JT}=
  (-1)^{\tilde{l}+s-\tilde{J}+\Lambda'+\tilde{\Lambda'}}\hat{j_{1}'^{2}}\hat{j_{2}'^{2}}
{\hat{\tilde{\Lambda'^2}}} {\hat{\Lambda'^{2}}}
\hat{\tilde{J}^{2}}\hat{J}^{2}\frac{2}{\pi} {\hat l}{\hat L}^\prime
(1-(-1)^{l^\prime +s^\prime +T})(1-(-1)^{{\tilde l}^\prime+{\tilde
s}^\prime+T})\hat{T^{2}}
$$
$$
a_{n^{\prime }l^{\prime }
 N^{\prime}L^{\prime }}^{n_1^{\prime }l_1^{\prime }n_2^{\prime
}l_2^{\prime }\Lambda ^{\prime }}a_{{\tilde n}^{\prime }{\tilde l}^{\prime }
 {\tilde N}^{\prime}{\tilde L}^{\prime }}^{n_1^{\prime }l_1^{\prime }n_2^{\prime
}l_2^{\prime }{\tilde \Lambda} ^{\prime }}$$
$$
 N_{j_{1}^\prime
j_{2}^\prime l_{1}^\prime l_{2}^\prime s_{1}^\prime s_{2}^\prime} (\lambda,
l,s,J,\tilde{l}\tilde{J};  s^\prime l^\prime \Lambda^\prime L^\prime; {\tilde
s}^\prime {\tilde l}^\prime
 {\tilde \Lambda}^\prime{\tilde
L}^\prime )
$$
and
$$
N_{j_{1}^\prime j_{2}^\prime l_{1}^\prime l_{2}^\prime s_{1}^\prime s_{2}^\prime}
(\lambda, l,s,J,\tilde{l}\tilde{J};  s^\prime l^\prime \Lambda^\prime L^\prime;
{\tilde s}^\prime {\tilde l}^\prime
 {\tilde \Lambda}^\prime{\tilde
L}^\prime )= \sum
_{x}(-1)^{\lambda+x+L'}(2\lambda+1)(2x+1)C^{\tilde{L'}0}_{L'0\lambda 0}
C^{\tilde{l}0}_{l0\lambda 0}\{j_{1}^{\prime }j_{2}^{\prime }x\}$$
$$\left \{
\begin{array}{c}
\lambda \tilde{J}J \\
sl\tilde{l}
\end{array}
\right \}
 \left \{
 \begin{array}{c}

 J{\tilde J}\lambda
\\
\tilde{L'}L'x
\end{array}
\right \} $$
\bigskip
$$\left\{
\begin{array}{c}
L^{\prime}xJ \\
s^{\prime }l^{\prime }\Lambda ^{\prime }
\end{array}
\right\} \left\{
\begin{array}{c}
{\tilde L}^{\prime }x{\tilde J} \\
{\tilde s}^{\prime }{\tilde l}^{\prime } {\tilde \Lambda }^{\prime }
\end{array}
\right\} \left\{
\begin{array}{c}
\begin{array}{c}
j_{1}^{\prime }xj_{2}^{\prime } \\
l_{1}^{\prime }\Lambda ^{\prime }l_{2}^{\prime }
\end{array}
\\
s_{1}^{\prime }s^{\prime }s_{2}^{\prime }
\end{array}
\right\} \left\{
\begin{array}{c}
\begin{array}{c}
j_{1}^{\prime }xj_{2}^{\prime } \\
l_{1}^{\prime }{\tilde \Lambda }^{\prime }l_{2}^{\prime }
\end{array}
\\
s_{1}^{\prime }{s}^{\prime }s_{2}^{\prime }
\end{array}%
\right\}
 $$
 Here,
  $\overline {\alpha} _{1} ^ {\prime}= n _ i ^ {^ {\prime}} l _ i ^ {^ {\prime}}
   j _ i ^ {\prime}t _ i ^ {^ {\prime}}$.
  The limitations on the angular momentums following from the 6j- symbols are
$(J\tilde{J}\lambda),(\lambda\tilde{l}l),(\tilde{J}\tilde{l}s),(sJl),
(J\lambda\tilde{J}), (\tilde{J}\tilde{L'}x), (\lambda\tilde{L'}L'),(L'Jx),(L'xJ),
(Js'l'),(xs'\Lambda '),(\Lambda 'L'l'), $ $(\tilde{L'}x\tilde{J}),
(\tilde{J}\tilde{s'}\tilde{l'}), (x\tilde{s'}\tilde{\Lambda '}),(\Lambda
'\tilde{L'}\tilde{l'})$.

  The limitations on the angular momenta from the 9j- symbols are:
$(j_{1}'xj_{2}'),(j_{1}'l_{1}'s_{1}'),(l_{1}'\Lambda'l_{2}'),(s_{1}'s's_{2}'),
(x\Lambda 's'),(j_{2}'l_{2}'s_{2}'), (l_{1}'\tilde{\Lambda '}l_{2}'),
(s_{1}'\tilde{s'}s_{2}'),$ $  (j_{1}'l_{1}'s_{1}'),(x\tilde{\Lambda '}\tilde{s '})$.
The limitations on the angular momenta following from the transition potential
coincide with limitations in the case of the ls- coupling.

\textbf{ References }\\
1. A. Kerman and L. Kisslinger, Phys. Rev. \textbf{180}, 1483 (1969). \\
2. H. Arenh$\ddot{o}$vel and M. Danos, Phys. Rev. Lett. \textbf{28B}, 299 (1968).\\
3. L. Kisslinger, Phys. Lett. \textbf{29B}, 211 (1969).\\
4. H. Arenh$\ddot{o}$vel, M. Danos and H. T. Williams, Phys. Lett. \textbf{31B},
109 (1970).\\
5. H. J. Weber and H. Arenh$\ddot{o}$vel, Phys. Rept. \textbf{C36}, 277 (1978).\\
6 . A. I. Amelin, M. N. Behr, b. A. Chernyshov et al., Phys. Lett.
\textbf{B337}, 261 (1994).\\
7. C. I. Morris, J. D. Zumbro, J. A. McGill et al., Phys. Lett. \textbf{B419}, 25
(1998). E. A. Pasyuk, R. L. Boudrie, P. A. M. Gram et al.,nucl-ex/9912004.\\
8. M. Liang, D. Brenford, T. Davinson et.al., Phys. Lett. \textbf{B411}, 244(1997).\\
9. G. M. Huber, G. J. Lolos, E. J. Brash et al., The TAGX Collaboration,
nucl-ex/9912001.\\
10. V. M. Bystritsky, A. I. Fix, I. V. Glavanakov et al., Nucl.
Phys. \textbf{A705}, 55 (2002).\\
11. G. Horlacher, H. Arenh$\ddot{o}$vel, Nucl. Phys. \textbf{A300},348 (1978).\\
12. H. Arenh$\ddot{o}$vel, M. Danos and H. T. Williams, Nuclear Phys. \textbf{A162},
12(1971).\\
13. D. A. Varshalovich, A. N. Moskalev and V. K. Khersonskii, Quantum Theory of
Angular Momentum, " Nauka"  Press, Leningrad (1975).\\
14.Gomez Tejedor, E. Oset, Nucl.Phys. \textbf{A571},667 (1994). \\
15. G. Horlacher, Doktorarbeit, Mainz, 1977.
\newpage
\begin{center}
{\textbf{Figures }}
 \end{center}

  \begin {figure}[!ht]
\unitlength = 1cm
 \centering
\includegraphics [width = 12cm , keepaspectratio] {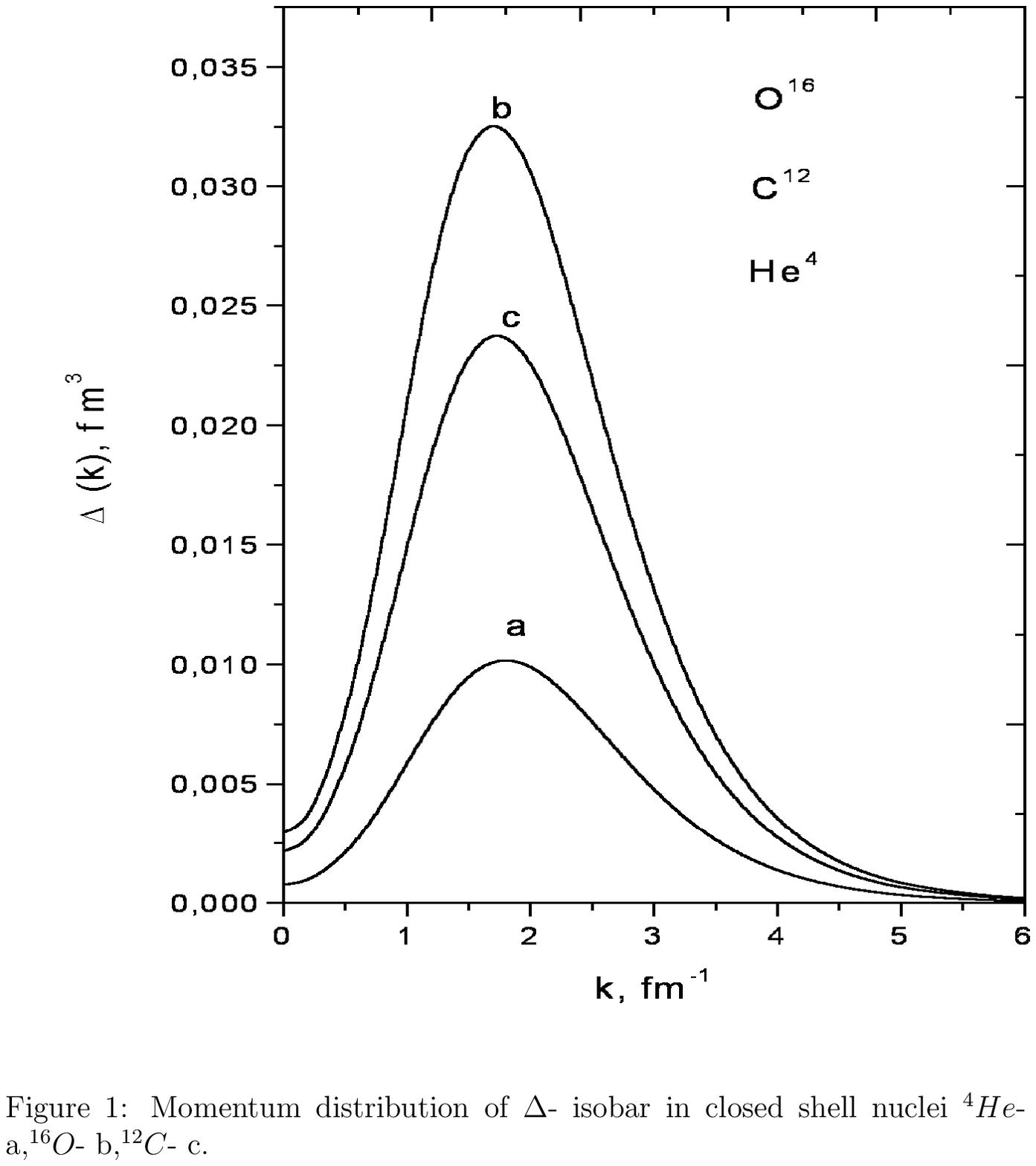}
\end {figure}

   \end{document}